# Experimental observation of topological exciton-polaritons in transition metal dichalcogenide monolayers

Mengyao Li[1,2,3*], Ivan Sinev[4*], Fedor Benimetskiy[4], Tatyana Ivanova[4], Ekaterina Khestanova[4], Svetlana Kiriushechkina[1], Anton Vakulenko[1], Sriram Guddala[1,2], Maurice Skolnick[4,5], Vinod Menon[2,3], Dmitry Krizhanovskii[4,5], Andrea Alù[6,3,1], Anton Samusev[4], Alexander B. Khanikaev[1,2,3]

[1]Department of Electrical Engineering, City College of New York, New York, NY, USA

[2]Physics Department, City College of New York, New York, NY, USA

[3]Physics Program, Graduate Center of the City University of New York, New York, NY, USA

[4]Department of Physics and Engineering, ITMO University, Saint Petersburg, Russia

[5]Department of Physics and Astronomy, University of Sheffield, Sheffield S3 7RH, UK

[6]Photonics Initiative, Advanced Science Research Center, City University of New York, New York, NY, USA

*These authors contributed equally to the present work

The rise of quantum science and technologies motivates photonics research to seek new platforms with strong light-matter interactions to facilitate quantum behaviors at moderate light intensities.[1–3] One promising platform to reach such strong light-matter interacting regimes is offered by polaritonic metasurfaces, which represent ultrathin artificial media structured on nano-scale and designed to support polaritons – half-light half-matter quasiparticles. Topological polaritons, or "topolaritons", offer an ideal platform in this context, with unique properties stemming from topological phases of light strongly coupled with matter. Here we explore polaritonic metasurfaces based on 2D transition metal dichalcogenides (TMDs) supporting in-plane polarized exciton resonances as a promising platform for topological polaritonics. We enable a spin-Hall[4,5] topolaritonic phase by strongly coupling valley polarized in-plane excitons in a TMD monolayer with a suitably engineered all-dielectric topological photonic metasurface. We first show that the strong coupling between topological photonic bands supported by the metasurface and excitonic bands in $MoSe_2$ yields an effective phase winding and transition to a topolaritonic spin-Hall state. We then experimentally realize such phenomena and confirm the presence of one-way spin-polarized edge topolaritons. Combined with the valley polarization in a $MoSe_2$ monolayer, the proposed system enables a new approach to engage the photonic angular momentum and valley degree of freedom in TMDs, offering a promising platform for photonic/solid-state interfaces for valleytronics and spintronics.

## Introduction

Topological photonics[6–8] has seen a tremendous progress in the past years with numerous topological phases implemented in a variety of platforms, from microwave to optical frequencies[9–20]. Enriching topological photonics by mixing light with condensed matter provides even more exciting avenues for controlling exotic states of light and matter. Indeed, integrating topological photonic systems with quantum wells and quantum dots has already led to major breakthroughs, such as topological lasers[21–24], topological polaritonic phases[25–28], active[29,30] and nonlinear[31–33] topological photonic devices. Consistent with their non-topological cousins, topolaritons[25,34,35] represent "half-light and half-matter" excitations emerging as the result of strong coupling[36–40] between electromagnetic and solid-state degrees of freedom. In addition, they are enriched by topological features. The combination of photonic topological properties (one-way spin-polarized transport, topological protection against scattering) and strong interactions arising from light-matter hybridization, may support phenomena such as topological solitons, modulation instability and generation of squeezed topological light[17,31,34,41–47]. Moreover, topolaritons pave the way towards the development of active topological nanophotonic devices with giant optical nonlinearity[48,49] enabling control of light by light at small intensities, down to the single photon level[50–53]. Overall, polaritonic systems serve as an ideal interface between photonics and solid-state systems, facilitating control of spin- and valley-degrees of freedom[54–63] in future quantum devices. Topological polaritons, enriched with additional degrees of freedom, inherited from nanoscale structured photonic materials, thus offer uniquely versatile control of quantum states with photons.

In this context topolaritons, have been an active subject of research with several recent theoretical and experimental findings, in particular in systems based on quantum wells integrated into photonic nanostructures[21,25–28,30]. However, topolaritonic systems reported so far have been mostly limited to 1D systems, and a 2D system characterized by 1D topological invariants[21,28]. The only topolaritonic system characterized by a 2D topological invariant, the Chern number, has been demonstrated recently by S. Klembt et al.[26] for the case of broken time-reversal symmetry. Specifically, 2D topological polaritons have been realized in Bragg micropillar (MC) lattices based on GaAs quantum wells (QW) under the application of a very strong external magnetic field. The realization of spin-Hall polariton, another 2D topolaritonic phase which does not require magnetic field, in structures based on GaAs QWs is rather difficult, since it requires etching through GaAs QW, leading to strong exciton surface recombination preventing polariton formation and making this platform rather challenging for topolaritonics.

In this work we put forward a new path to spin-Hall topolaritonics, which does not require magnetic fields and it is based on a versatile platform of polaritonic metasurfaces containing monolayer transition metal dichalcogenides (TMDs). Our approach leverages the large exciton dipole moment in a monolayer semiconductor and the remarkable compatibility of 2D materials with various photonic structures to realize strong coupling regime between light and matter. We show that topological spin-Hall photonic metasurface integrating a TMD monolayer with a pair of degenerate valley-polarized excitons in the strong coupling regime gives rise to topological transitions and the formation of a topolaritonic phase characterized by nonvanishing spin-Chern number. Introduction of domain walls separating topological and trivial phases is then shown to give rise to the spin-polarized topolaritonic boundary modes. Spin-locking of these boundary modes and their selective coupling to circularly polarized light of opposite handedness enables unique polaritonic spin-Hall phenomena.

The broadband spin-Hall topolaritons are experimentally realized by integrating unpatterned $MoSe_2$ monolayer into spin-Hall type photonic structures. Very large exciton dipole moments in $MoSe_2$ monolayers lead to the formation of bulk topolaritons with large Rabi splitting. We demonstrate that topological domain walls of such structures support edge polaritons exhibiting spin-polarized one-way propagation, and whose excitonic components are valley polarized due to the valley polarization in $MoSe_2$. Thus, we demonstrate for the first time 2D topolaritonic phase which does not require magnetic field, exhibits large topological gap and broadband one-way topolaritonic boundary modes. We propose our platform for coupling the angular momentum of photons to the valley degree of freedom of excitons in transition metal dichalcogenides, which can be employed as a resilient topological interface between photonic and electronic components in future valleytronic devices.

## Topolaritonic metasurface

The structure we consider here is schematically shown in Fig. 1**a** and represents a Si photonic metasurface supporting photonic topological spin-Hall-like phase[64–66] with $MoSe_2$ monolayer placed on top of it. The metasurface represents a honeycomb shrink-expand lattice design is based on the one proposed for a topological quantum optical interface[17,67], which we adjusted to hold semi-leaky (trapped by the metasurface, but coupled to the radiative continuum) topological modes near the exciton frequency in $MoSe_2$. The semi-leaky character of the modes allows us to directly excite and probe both photonic and polaritonic modes supported by such a metasurface[68,69]. The $MoSe_2$ monolayer is placed on top of the metasurface together with a thin hBN layer. Apart from the general purpose of enhancing the quality of the exciton in the monolayer, the hBN layer enables fine tuning of the parameters of our system. Firstly, if placed between the metasurface and the monolayer, it allows the careful control of the coupling strength between photonic and exitonic subsystems by changing its thickness. Secondly, an increase of hBN thickness also gradually red-shifts the photonic modes of the metasurface, which allows to arrange the position of the topological gap with respect to the exciton (see Supplement G for experiments with thick hBN layers). In addition, the possibility to change the spectral position of the exciton resonance by adjusting temperature enabled spectral tuning of the crossing between the photonic and excitonic states, which was used to explore different topological regimes.

This type of photonic metasurface is known to support a photonic analogue of the spin-Hall phase protected by the lattice symmetry. The topological transition in this system occurs when the lattice is distorted from an unperturbed graphene-like lattice of triangular holes, in which case the system has a characteristic dispersion with two overlapping Dirac cones, shown in Fig. 1**b** (the black lines). The pairs of upper and lower cones correspond to the clockwise and counterclockwise $s = \pm 1$ circularly polarized dipolar $p_\pm = p_x \pm i p_y$ $(l = \pm 1 = 1 \times s)$ and quadrupolar $d_\pm = d_{xy} \pm i d_{x^2-y^2}$ $(l = \pm 2 = 2 \times s)$ bands.

In order to induce a topological transition in this lattice[64,67,68], its symmetry is reduced by shrinking (Fig. 1**a**, blue unit cells) or expanding the six nearest triangular holes (Fig. 1**a**, red unit cells), rendering a triangular lattice of hexamers. As seen in Fig. 1**b**, such symmetry reductions are accompanied by the opening of a photonic band gap in place of the Dirac cones, which, for the expanded lattice scenario, yields the inversion of dipolar and quadrupolar bands leading to a photonic topological spin-Hall phase. Numerous experimental studies of topological photonic systems based on this scheme unambiguously confirmed a robust spin-polarized topological boundary transport in microwave and optical domains.[17,68,70,71]

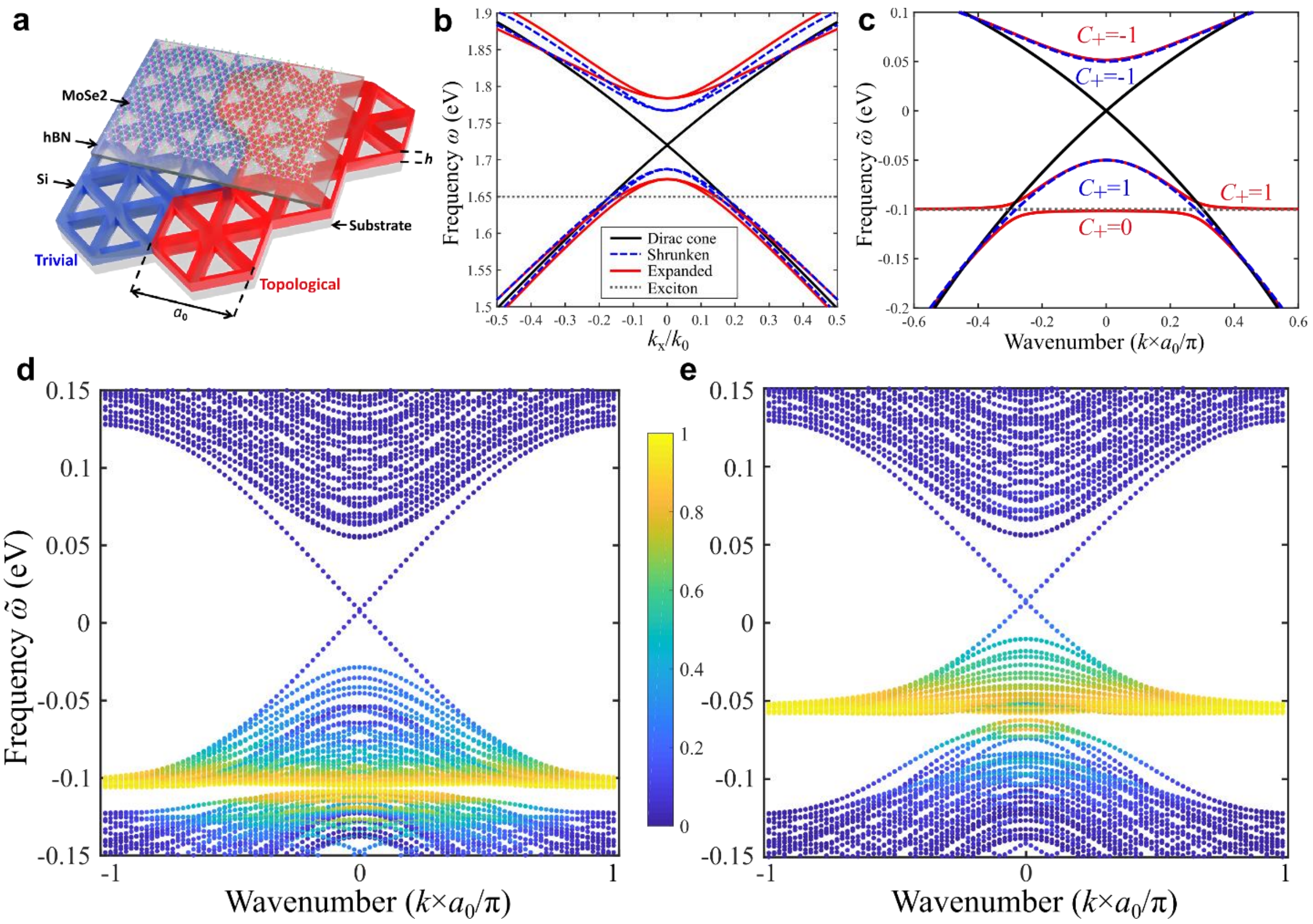

**Fig. 1| Topolaritonic metasurface integrating transition metal dichalcogenide monolayer**. **a**, schematic image of topological metasurface with hBN spacer and $MoSe_2$ monolayer on top. Lattice constant is $a_0$=460nm, Si layer thickness is $h$=75nm. **b**, Numerically calculated photonic band structure for the cases of gapless (black lines), topological (red lines) and trivial (blue lines) metasurfaces. The position of the excitonic band is shown by the dashed horizontal line. **c**, Bulk band structures of topolaritonic system obtained from the analytical model Eq (1) for different values of the mass term $M$ and the exciton-photon coupling $q_d$. The black dashed line shows the case of no coupling ($q_{d(p)} = 0$) in the gapless case ($M = 0$) to reveal the Dirac cone and the flat excitonic band ($\widetilde{\omega}_{ex} = -1.5$ eV). The blue lines show the case of a topological photonic system without coupling to polaritonic subsystem ($q_{d(p)} = 0$) for $M = 1$. Red lines show the case of topolaritonic transition via strong coupling with $q_p = 2$ and $q_d = 0$. **d**, **e**, Topolaritonic spin-up boundary modes for the two scenarios of excitonic band within or at the edge of the lower (dipolar) photonic band. The two, forward and backward solutions, are spatially separated and appear at the two opposite domain walls of the supercell with the periodic boundary condition. The edge states are clearly band-crossing (and anti-crossing) between the excitonic and photonic bands and approach the former at lower frequencies, confirming its polaritonic nature. The degree of excitonic fraction of the bands shown in color is calculated as the ratio of the amplitude in the excitonic component of the eigenstate vector to the norm of the vector.

By adding a TMD monolayer on top of this photonic structure we introduce excitonic degrees of freedom. It is important to stress that excitons in $MoSe_2$ are (***i***) characterized by non-zero orbital momentum of $m = +1$ and $m = -1$ at K and K' valleys, respectively, and thus establish time-reversal partners essential for the topolaritonic spin-Hall phase engineered here, and (***ii***) polarized in the plane of the monolayer, which allows to efficiently couple the in-plane electric field of the modes supported by the metasurfaces to the excitons[69]. We note that the characteristic size of excitons (~ 1 nm) and the scale of field variations of the photonic modes are orders of magnitude apart, which implies that excitons can interact with both dipolar and quadrupolar photonic modes. Nonetheless, while the angular momentum of the two subsystems is not in correspondence, the photon spin (the handedness of both dipolar and quadrupolar modes) still is in one to one correspondence with the orbital momentum and the valley degree of freedom of excitons due to angular momentum conservation (see discussion in Supplement D). This can be explained simply as the result of the selection rule on the interactions of light and excitons, which are defined exclusively through the spin ($s = \pm 1$) degree of freedom of photonic modes. At the same time, the non-vanishing orbital momentum ($l$=1, 2) of photonic modes due to the lattice structure does not play any fundamental role in the interactions, but only defines the strength of coupling due to the nonuniformity of the field profiles (Supplement D). Nonetheless, this orbital momentum of photons is crucial for the formation of the topological photonic phase and, therefore, for the physics of our topolaritonic system. Indeed, in the photonic metasurface, the spin and orbital momentum of light are entangled with each other due to the spin-orbital coupling, leading to the topological behavior emulating quantum spin-Hall effect. This entanglement enables an indirect coupling of the orbital momentum of light and that of excitons, which facilitates the topological polaritonic phase reported here.

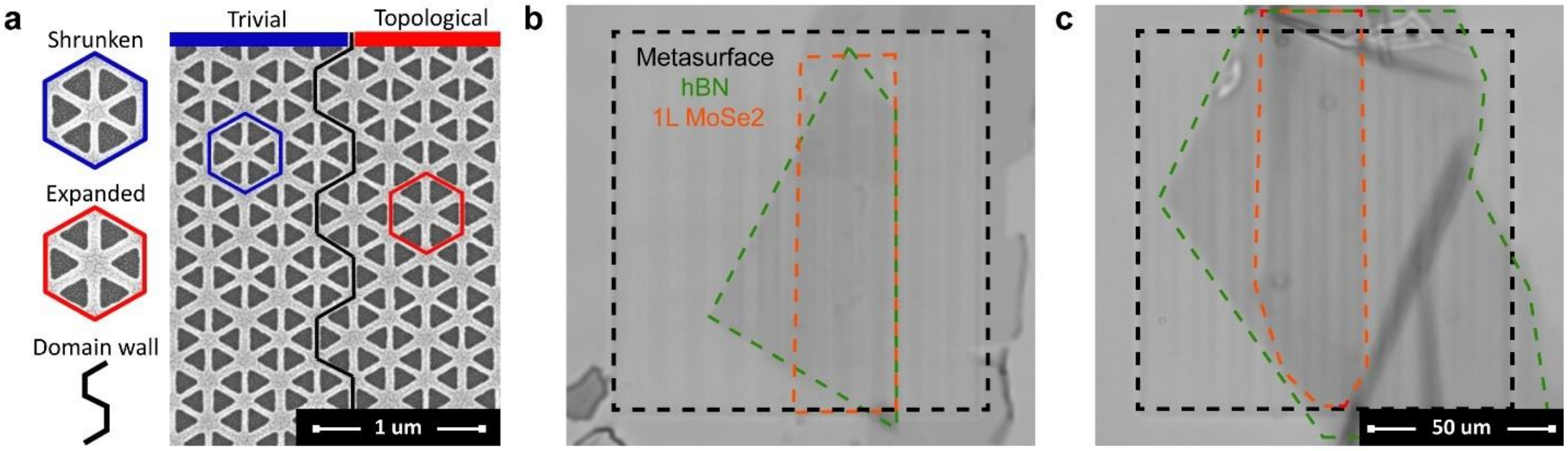

**Fig. 2| Experimental samples of topolaritonic metasurfaces. a**, SEM images of topological photonic metasurface with unit cells of trivial and topological domains indicated by the hexagon and the domain wall shown by the armchair shaped line. **b** and **c**, optical microscope images of the two topolaritonic metasurface (black) samples with $MoSe_2$ monolayers (orange) and 12 nm thick hBN spacers (green). **b**, $MoSe_2$ monolayer transferred directly onto the metasurface and covered with an hBN flake. **c**, monolayer transferred on top of hBN flake.

## Theoretical description

The topolaritonic system under study can be described by an effective Hamiltonian of the form

$$\hat{\mathcal{H}} = \hat{\mathcal{H}}_{ph} + \hat{\mathcal{H}}_{ex} + \sum_{\boldsymbol{k},l,m,s} \left( g_{(\boldsymbol{k},l,m,s)} \hat{b}_m^{+} \hat{a}_{\boldsymbol{k},l,s} \delta_{s,m} + g^{+}_{(\boldsymbol{k},l,m,s)} \hat{b}_m \hat{a}^{+}_{\boldsymbol{k},l,s} \delta_{s,m} \right), \tag{1a}$$

$$\hat{\mathcal{H}}_{ph} = \sum_{\boldsymbol{k},l,l',s,s'} \hat{H}_{(\boldsymbol{k},l,l',s,s')} \hat{a}^{+}_{\boldsymbol{k},l',s'} \hat{a}_{\boldsymbol{k},l,s}, \tag{1b}$$

$$\hat{\mathcal{H}}_{ex} = \sum_{m,m'} \tilde{\omega}_{ex} \hat{b}^{+}_{m'} \hat{b}_m, \tag{1c}$$

where the photonic Hamiltonian $\hat{\mathcal{H}}_{ph}$ is obtained from electromagnetic theory (detailed in Supplement B) and it assumes the form of the well-known Bernevig-Hughes-Zhang (BHZ) Hamiltonian[5] of the two-dimensional $Z_2$ topological insulator

$$\hat{\mathcal{H}}_k = B_2|\mathbf{k}|^2 \hat{I} + \begin{pmatrix} M - B|\mathbf{k}|^2 & A(-ik_x + k_y) & 0 & 0 \\ A(ik_x + k_y) & -M + B|\mathbf{k}|^2 & 0 & 0 \\ 0 & 0 & M - B|\mathbf{k}|^2 & A(-ik_x - k_y) \\ 0 & 0 & A(ik_x - k_y) & -M + B|\mathbf{k}|^2 \end{pmatrix}. \tag{2}$$

Here, an energy shift $\omega_0$, equal to the frequency of the Dirac point for the unperturbed lattice, was introduced so that the Dirac point arises at zero energy. In what follows we express frequency in electron-volt units. In Eqs. (1) $\hat{a}_{\boldsymbol{k},l}$ and $\hat{b}_{\boldsymbol{k},l}$ ($\hat{a}^{+}_{\boldsymbol{k},l}$ and $\hat{b}^{+}_{\boldsymbol{k},l}$) are the annihilation (creation) operators for photons and excitons, respectively, and the label $\boldsymbol{k}$ corresponds to the Bloch momentum of photonic modes, $\tilde{\omega}_{ex} = \omega_{ex} - \omega_0$ is the exciton frequency shifted by $\omega_0$, and $g_{(\boldsymbol{k},l,m,s)}$ describes the coupling between photonic and excitonic degrees of freedom in our system. We note that the subscript $m = \pm 1$ simultaneously describes the orbital momentum of excitons and their valley degree of freedom (K or K') due to the valley polarization in the TMD monolayer. As described in Supplements A and B, the 4x4 structure of the photonic Hamiltonian Eq. (2) incorporates both spin $s$ and orbital momentum $l$ degrees of freedom of photonic modes, while the mass term $M$ reflects the band inversion.

The exciton-photon coupling is crucial for generating topolaritons [4]. In our case, this coupling, described by the 2x4 matrix

$$g_{(\boldsymbol{k},l,m,s)} = \begin{pmatrix} q_{p_+} & q_{d_+} & 0 & 0 \\ 0 & 0 & q_{p_-} & q_{d_-} \end{pmatrix}, \tag{3}$$

ensures that angular momentum is conserved, which is reflected by the $\delta_{s,m}$ factors in Eq. (1), and $q_{p_+} = q_{p_-}$, $q_{d_+} = q_{d-}$ when the time-reversal symmetry is preserved. This ensures that the spin-orbital coupling in the original photonic system is transferred to the exciton-polaritons once the hybrid system is in the strong coupling regime. As shown below, and detailed in Supplement E, the nonvanishing Berry flux in the topological photonic system introduces the (Berry) phase winding into the interaction of photons and excitons. However, as opposed to the case of the original proposal by Karzig et al.[25], our system does not break the time-reversal symmetry, which is responsible for the presence of the two time-reversal partners for both photonic ($s = \pm 1$) and excitonic ($m = \pm 1$) subsystems. Along with the valley polarized character of the exciton-photon coupling ($s = m$), this enables a completely new spin-Hall topolaritonic phase.

To show the emergence of winding responsible for the topolaritonic phase, we first focus on the upper left (spin-up $s = 1$) block of the photonic Hamiltonian Eq. (2) and note that the respective eigenstates have a (typical for Dirac systems) form $|\psi_{s=1}\rangle \sim [1, f_{1(2)}\exp(\pm i\theta_{\boldsymbol{k}}), 0,0]^T$, where the $f_{1(2)}$ are the real-valued factors and the subscript 1(2) and the plus (minus) sign in the exponent, correspond to the lower (upper) spin-up bands, respectively. Since the exciton-photon interactions do not break the spin-degree of freedom, it makes it legitimate to focus only on spin-up 2x2 block of the photonic Hamiltonian (2). Then, the respective Hamiltonian can be diagonalized by the unitary transformation $\widehat{U} \sim [1, 1; f_1\exp(i\theta_{\boldsymbol{k}}), f_2\exp(-i\theta_{\boldsymbol{k}})]$. For simplicity, we will assume that K-valley ($m = 1$) exciton interacts only with the dipolar band ($q_{p_+} \neq 0$, $q_{d_+} = 0$), which, as seen from the experimental results, is good approximation when the exciton frequency crosses only with that band (also see Supplement E for the theoretical justification). By expanding the transformation $\widehat{U}$ onto larger 3x3 Hilbert space to span the excitonic degree of freedom, we obtain a new 3x3 unitary operator $\widehat{\widehat{U}} = diag[\widehat{U},\ 1]$. This new transformation diagonalizes the spin-up photonic 2x2 block, it does not affect the exciton frequency, but it changes the coupling by adding an extra phase $\tilde{g}_+ = \widehat{U}[q_{p_+}, 0] = [q_{p_+}, \quad q_{p_+}\exp(-i\theta_{\boldsymbol{k}})]$, which clearly shows the emergence of winding similar to that in topolaritons with broken TR symmetry[25]. In our case, however, because TR symmetry is not broken, we have another (spin-down) time-reversal topolaritonic partner which experiences coupling with the opposite phase winding $\tilde{g}_- = [q_{p_-} \quad q_{p_-}\exp(i\theta_{\boldsymbol{k}})]$. Our calculations indeed confirm that this winding leads to the nonzero value of the spin-Chern numbers of $+1$ and $-1$ for the spin-up and spin-down polaritons, respectively. However, if the band crossing takes place with the upper (quadrupolar) band ($q_{d_+} \neq 0$), it gives rise to the opposite windings for the same spins, implying the reversal of signs of the spin-Chern numbers for spin-up and spin-down bulk topolaritons.

In Fig. 1**c** we show the band structure calculated from the effective Hamiltonian Eq. (1) for the case of the excitons crossing lower (dipolar) photonic bands. For the case of the expanded (topological) photonic lattice, without coupling to excitons, the photonic bands are found to have the spin-Chern numbers $C_{ph} = \pm 1$ for the two lower bands and $C_{ph} = \mp 1$ for the two upper bands of opposite spins. In agreement with our predictions, turning on the exciton-photon coupling leads to the transition to the topolaritonic state. We observe that the strong coupling gives rise to the avoided crossing of excitonic and photonic bands, while calculations of the spin-Chern numbers for the excitonic bands yield nonzero values identical to those of the crossed photonic bands, i.e. $C_{ex} = \pm 1$.

Below we focus on the experimental realization of two scenarios, when crossing of the excitons occurs with the lower (dipolar) photonic band (***i***) inside the band, and (***ii***) at the edge of the band. While in both cases the strong coupling leads to avoided crossing of excitonic and photonic bands, and to the formation of topolaritonic phase, the degree of excitonic fraction in the resultant edge states is different.

Band structure calculations for the first scenario (obtained with the tight-binding model (TBM) as described in Supplement A) are shown in Fig. 1**d**, where the degree of the excitonic component of the modes is encoded in the color of the bands. As expected for this scenario, the flat section of the second band (the remnant of the excitonic flat bands) is close to 100% excitonic, but excitonic fraction fades away as we move into the parabolic portion of the band (the reminiscent of the dipolar photonic band). Since for the original topological photonic system the major contribution to the spin-Chern number stems from the Berry curvature near the tip of this parabolic band in the proximity of Γ point, it is not surprising that it is carried over to the polaritonic band after the interaction with excitons is introduced. Indeed, a direct calculation of the spin-Chern number confirms that the topological charge is transferred from the photonic band to the polaritonic band due to the strong coupling and the avoided crossing. As a consequence, the topolaritonic boundary modes resemble those of the original photonic system; they are crossing the stop band, but also experience an additional bending towards the flat portion of the bulk polaritonic bands. It is near this flat portion where their polaritonic content reaches maximal values, while the modes appear to be almost completely photonic in the middle and at the upper edge of the topological bandgap.

For the second scenario, the excitonic bands touch the tip of the edge of the lower (dipolar) topological photonic bands at the Γ point of the photonic Brillouin zone. Exciton-photon interactions for this case again give rise to the avoided crossing and slight upward and downward curving of the bulk bands. This special situation allows for an exact analytical treatment which shows that the both bands appear to be ~50% excitonic. The phase winding in the interaction between excitons and photons in this case renders the polaritonic band topological, while the photonic band loses its topological properties (Supplement E). Remarkably, this scenario also yields topological boundary states with large excitonic components. Indeed, as can be seen from Fig. 1**e**, the topolaritonic boundary modes appear to be highly excitonic in a wide range of wavenumbers and energies below the mid-gap frequency, and even have a significant excitonic fraction in the mid-gap and at higher frequencies.

## Experimental results

The design of our topological photonic metasurface was optimized to exhibit topological band crossing in the vicinity of the exciton resonance in $MoSe_2$ and to facilitate tuning of the crossing frequency by the temperature. The resultant band structures for optimized shrunken (trivial) and expanded (topological) metasurfaces, calculated in COMSOL Multiphysics finite element method (FEM) solver, are shown in Fig. 1**b**. The final designs were fabricated by patterning Silicon on Insulator (SOI) substrates with the use of e-beam lithography followed by reactive ion etching. The details of the fabrication techniques used can be found in Methods. The fabricated samples consist of multiple shrunken and expanded regions forming an array armchair-shaped domain walls[68]. The bulk regions of 10 periods were confirmed to be wide enough to eliminate possible coupling between the edge states confined at the domain walls. The proximity of one of such the domain walls separating topological and trivial regions is shown in an SEM image Fig. 2**a**.

While multiple samples were fabricated, here we describe two samples corresponding to the two scenarios described in the theoretical section. The selected samples were picked making sure that the exciton resonance in $MoSe_2$ ($\omega \approx 1.65$ eV at 7K) would cross the lower-frequency photonic bands, either through or near the edge of the band at the Γ-point of the photonic band structure. Optical microscope images of the respective samples are shown in Figs. 2**b** and **c**, and correspond respectively to (***i***) the case of the $MoSe_2$ monolayer on top of the metasurface with subsequent transfer of a 12 nm hBN layer and (***ii***) the case of a 12 nm hBN spacer placed on the metasurface with subsequent transfer of the $MoSe_2$ monolayer on top of it. The boundaries of both $MoSe_2$ monolayers and hBN flakes on top of the metasurfaces are indicated by color lines in Figs. 2**b**,**c**. We note that the small thickness of hBN monolayer did not affect the coupling strength in the two scenarios, unlike in other cases with thicker hBN (Supplement G). However, due to relatively high refractive index of hBN, even a 12 nm layer led to a noticeable spectral redshift of the photonic bands, which we employed for high-precision control of the position of topological gap with respect to the exciton.

The semi-leaky character of the photonic and polaritonic bands allows their optical characterization by the back focal (Fourier) plane imaging in our custom-built experimental setup, which enables extraction of the band diagrams in frequency-momentum space at cryogenic temperatures (see Methods for details on the experimental techniques used).

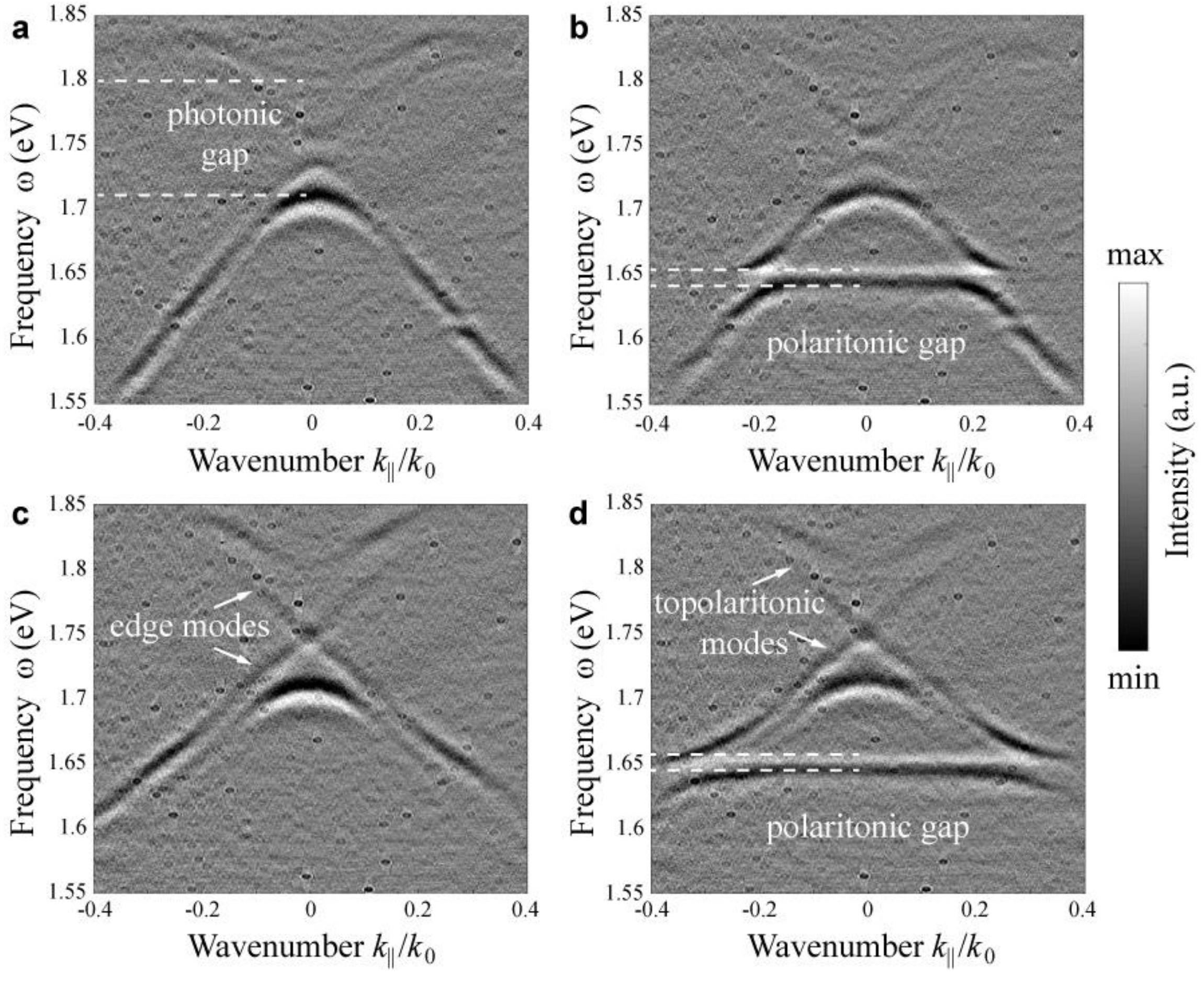

**Fig. 3| Formation of topolaritonic bands and edge states after the introduction of MoSe2 exciton**. **a** and **c**, angle-resolved reflectance measured from the vicinity of the domain wall of a metasurface without $MoSe_2$ for TE and TM polarization, respectively, which reveal predominantly bulk and edge spectra,

respectively. **b** and **d**, reflectance maps after the transfer of 12 nm hBN spacer and a MoSe2 monolayer on top of the metasurface demonstrating strong coupling topolaritonic regime and the respective bulk and edge topolaritonic bands.

We first studied the sample without the Van der Waals heterostructure. As seen in Fig. 3**a,** which corresponds to the case of dominating bulk spectrum, the photonic lattice indeed exhibits a photonic band gap (~1.7-1.8 eV), while Fig. 3**b** reveals two gapless modes inside the bulk bandgap, which correspond to the spin-up and spin-down topological photonic edge states propagating along the domain wall in the opposite directions. The transfer of a 12 nm hBN flake followed by a $MoSe_2$ monolayer (Fig. 3**c**,**d**) to the structure leads to the appearance of the flat excitonic band crossing the lower photonic band of the topological metasurface approximately 0.07 eV below the band edge. These experimental spectra clearly reveal the anti-crossing behavior between excitonic and photonic bands with Rabi frequency of $\Omega_R = 0.02$ (obtained by fitting the data to the theoretical model), manifesting the strong coupling regime and the formation of topological bulk states, and, more importantly, of the formation of the topological edge polaritons. As an important indication of the polaritonic nature of the edge states in Fig. 3**d**, we notice that the respective bands asymptotically approach the polaritonic bulk band. In agreement with the bulk boundary correspondence[4,72], this result confirms that the topological invariants, i.e., the spin-Chern numbers, were transferred to the respective upper polaritonic bands from the former photonic band due to the strong coupling regime. Accordingly, the lack of edge states within the bandgap between upper and lower polaritons (former excitonic and photonic bands) further evidences that the spin-Chern number of the lower band is zero due to this transfer.

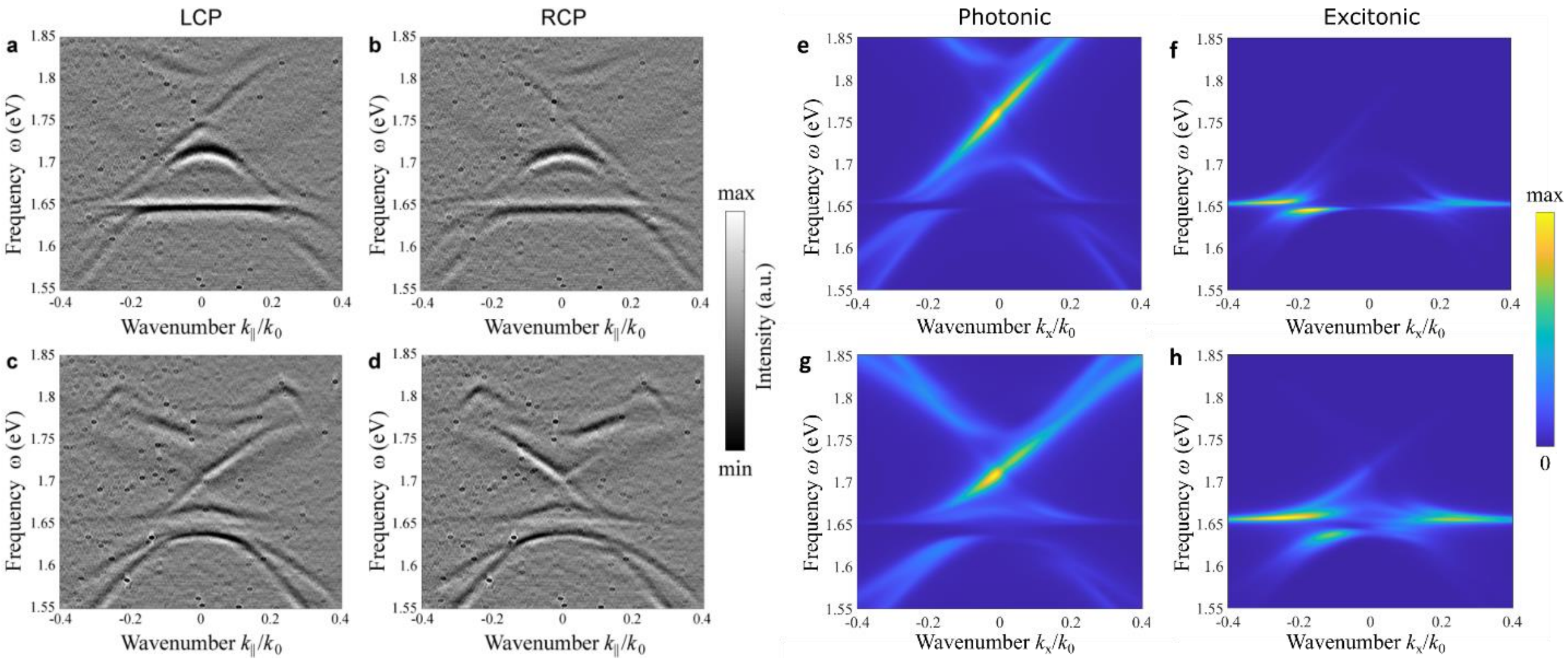

**Fig. 4| One-way propagation of circularly polarized edge topolaritons along the domain walls.** Two topolaritonic metasurfaces integrated with $MoSe_2$: with exciton deep inside the photonic bulk mode (top row) and at its edge (bottom row). **a-d** Experimental images for the case of exciton frequency deep in bulk and at edge with different circular polarizations between **a** and **b**, as well as **c** and **d**. **e-h** Corresponding case in TBM/CMT excitation model with fitting parameters emulating experiments. **e** and **g** are the photonic part of excited states, while **f** and **h** are excitonic components, for the two cases.

Perhaps the most valuable property of spin-Hall topological systems is the one-way spin-polarized character of their topological boundary states. While this property was observed experimentally in photonic structures, a similar property should also emerge in spin-polarized topolaritonic boundary modes. However, in our case the valley-polarization in the TMD monolayer, along with the selective interaction of photons of particular handedness with the excitons at a particular valley, provides an opportunity to entangle the photonic spin and orbital degrees of freedom and valley degree of freedom in $MoSe_2$. Therefore, in addition to the possibility of directional (one-way) excitation of edge topolaritons by circularly polarized light, we ensure that their excitonic component in the TMD monolayer is also valley polarized. Therefore, as a next step we collected the back focal plane reflectance spectra with left and right circularly polarized beams focused on the domain wall. The results depicted in Figs. 4**a**,**b** confirm the one-way nature of the edge topolaritons. The numerical model of the supercell with 10 topological and 10 trivial unit cells separated by the domain wall based on TBM/CMT modes was used to fit this structure with the results plotted in Fig. 4**e**,**f** which show photonic and excitonic components, respectively, in the vicinity (four unit cells) to the domain wall, excited by circularly polarized incident field. As expected, the one-way edge state has a strong excitonic component (Fig. 4**f**), which reaches nearly 100% near the exciton resonance and gradually fades away at higher energies where the photonic component (Fig. 4**e**) starts to dominate. Nonetheless, the excitonic component is retained and can be as high as 7% midgap, near the crossing between forward and backward edge states, for the case in Fig. 4

Next, we performed measurements on the second sample, with $MoSe_2$ placed right on top of the metasurface and hBN transferred on top to ensure similar dielectric environment as for the previous sample (the same spectral shift of photonic bands). The metasurface itself, however, was scaled accordingly to ensure that the exciton resonance is aligned with the very edge of the photonic band at the Γ-point of the photonic Brillouin zone. The corresponding measured band diagrams, shown in Figs. 4**c**,**d**, reveal that the upper polaritonic band resembles the excitonic flat band curving up due to the interaction between excitonic and photonic bands. The former photonic band, in turn, is pushed down, to yield the lower polariton. A complete polaritonic bandgap with Rabi frequency of $\Omega_R = 0.033$ eV separates these upper and lower bulk polaritons. As follows from our theoretical analysis using TBM/CMT model, in this case the avoided crossing should again yield one-way spin-polarized topolaritonic boundary states, but with even higher valley polarized excitonic component. The experimental results in Figs. 4**c**,**d** confirm that the bandgap hosts two counterpropagating spin-polarized one-way boundary states, with the forward and backward states that can be excited only with the circularly polarized light of proper helicity. Theoretical TBM/CMT model used to fit this scenario confirms the presence of one-way topolaritonic edge states with high degree of excitonic component, indicating that the edge states also carry a significant valley polarization along with the photonic spin, and polaritonic component can be as high as 20% midgap at the crossing between forward and backward edge states.

Finally, we showed tuning of the topolaritonic regime by shifting the frequency of the exciton resonance by changing the ambient temperature. The corresponding results in Fig. 5 show the cases of excitation with linearly polarized source at T=7K, 100K, and 200K, which results in the change of the spectral position of exciton with respect to the topological band edge affecting the degree of excitonic component of the edge state. Besides the possibility of tuning, it is notable that, even at relatively high temperatures, the excitons and photons exhibit strong coupling leading to the emergence of a topolaritonic quantum spin-Hall-like phase, hinting to the possibility of controlling the degree of valley polarization of topological exciton-polaritons by temperature, of great interest for valleytronic applications.

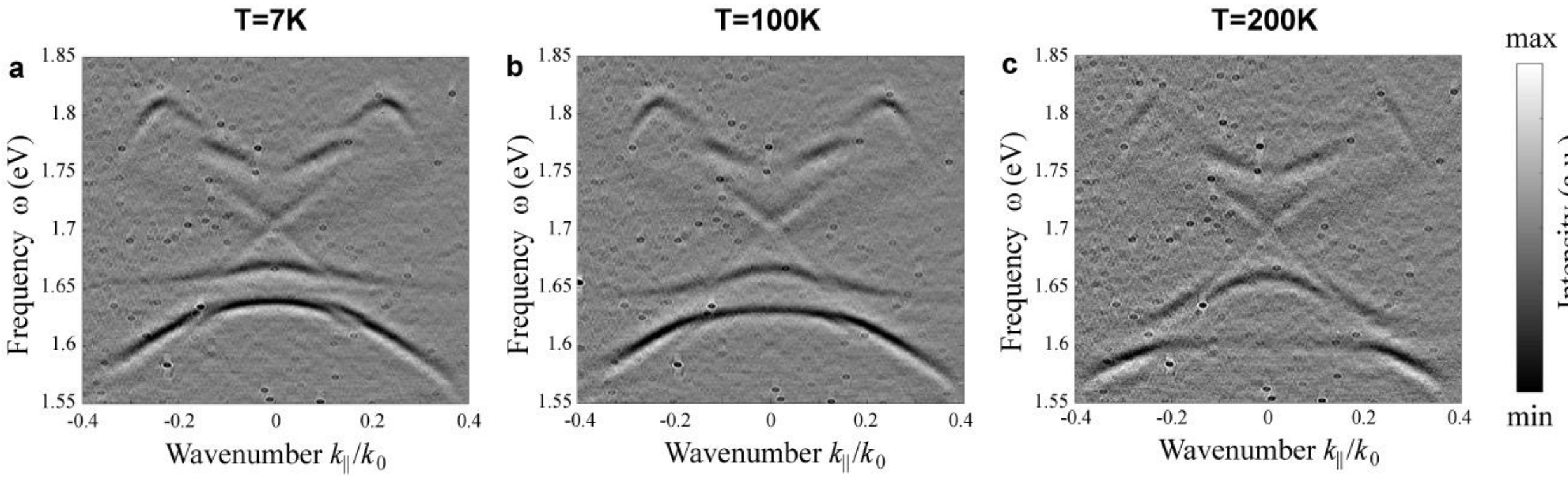

**Fig. 5| Tuning topolaritons dispersion through temperature control over the exciton frequency**. The dispersion of the bulk and edge states extracted from the back focal plane TM polarized reflectance maps at **a** T=7K, corresponding to the exciton frequency of ~1.65 eV, **b** T=100K, corresponding to the exciton frequency of ~1.63 eV, and **c** T=100K, corresponding to the exciton frequency of ~1.61 eV.

## Conclusions

To summarize, here we have introduced a new approach to engineer the topological polaritonic phases with preserved time-reversal symmetry by strongly coupling topological photonic systems with exitonic degrees of freedom in 2D materials. The strong in-plane polarizability and, more importantly, the presence of two time-reversal partner excitons at K and K' valleys in the 2D semiconductors, leads to strong coupling and avoided crossing behavior accompanied by the emergence of effective winding and topological transition to the topolaritonic spin-Hall phase. Time reversal symmetry and conservation of angular momentum in the system ensures that the excitons with opposite orbital momenta couple with photons of respective spins, thus ensuring the formation of two time-reversal partner topolaritonic bulk bands carrying nonzero spin-Chern numbers. This gives rise to the emergence of spin-polarized one-way edge topolaritons, which, at the same time, carry valley-polarized polaritonic component.

Our work paves the way to engineering novel topological phases in hybrid photonic-excitonic structures by enriching these systems with additional degrees of freedom inherited from their solid-state component, such as valley degree of freedom in TMDs. It has clear advantages over the conventional approach based on semiconductor heterostructures since 2D materials support a broad range of excitations with a variety of internal degrees of freedom and are easy to integrate into topological photonic systems. Thus, this concept can be extended to a wide range of solid-state systems hosting different excitation, e.g., phonons, polarons, magnons and spin-waves, which can be devised to interact with various topological photonic systems, e.g. regular and higher-order topological insulators, with and without time-reversal symmetry, yielding unprecedented topological phases and novel ways to control matter with light in a robust and a resilient manner. Strong and resilient light-matter interactions in such systems will facilitate enhanced nonlinear effects and novel quantum effects involving half-light and half-mater excitations, which can be of immense value for various classical and emerging quantum applications.

## Methods

### Sample fabrication:

A triangular lattice formed by the hexamers of triangular-shaped holes was fabricated on the Silicon-on-Insulator substrates (75 nm of Si, 2 μm of $SiO_2$) with the use of E-beam lithography (Elionix ELS-G100). First, the substrates were spin-coated with e-beam resist ZEP520A of approximately 170 nm thickness and then baked for 4 minutes at 180°C. Next, gold film 15nm thick was sputtered on top of resist. E-beam lithography exposure was followed by gold etch and the development process in n-Amyl Acetate cooled to 0°C for approximately 35 sec. Then, anisotropic plasma etching of silicon was conducted in The Oxford PlasmaPro System ICP by a recipe based on C4F8/SF6 gases. Triangular shaped holes were etched to the depth of about 75 nm at temperature 5°C etching with rate about 1.5 nm/sec. Finally, the residue of resist was removed by sample immersion into NMP heated to 60 °C.

A monolayer of $MoSe_2$ TMD material was exfoliated onto a thick PDMS stamp using standard tape technique and transferred to the substrate by the custom-built transfer stage. The monolayer was annealed at 350$^o$C for 2 hours to remove polymer residue from the transfer process. Further, some of the monolayers were encapsulated with a 12 nm hBN layer and annealed again at 350$^o$C for another 2 hours.

### Experimental set up:

Angle-resolved reflectance measurements were performed in a back focal plane configuration with a slit spectrometer coupled to a liquid-nitrogen-cooled imaging CCD camera (Princeton Instruments SP2500+PyLoN), using white light from a halogen lamp for illumination. The sample was mounted in an ultra-low-vibration closed-cycle helium cryostat (Advanced Research Systems) and maintained at a controllable temperature down to 7 K. To resolve the topological edge modes, we used a slit-type spatial filter in the image plane of the detection channel that allowed to collect signal only from the vicinity of a single domain wall. The measured angle-resolved reflectivity maps were post-processed to suppress the Fabry-Pérot background originating from the bottom silicon layer of the SOI substrate (see Supplement F).

### Data availability

Data that are not already included in the paper and/or in the Supplementary Information are available on request from the authors.

### Acknowledgements

The work was supported by the National Science Foundation with grants No. DMR-1809915 and NSF QII TAQS OMA-1936351, by the Defense Advanced Research Project Agency Nascent

Program, and by the Simons Foundation. The work on transferring TMD monolayers and hBN crystals on top of metasurfaces was supported by the Ministry of Science and Higher Education of Russian Federation, goszadanie no. 2019-1246 and megagrant 14.Y26.31.0015. Optical measurements were funded by RFBR, project number 20-32-70185. DNK and MSS acknowledge the support from UK EPSRC grants EP/R04385X/1 and EP/N031776/1.

## Author contributions

All authors contributed extensively to the work presented in this paper. ML and IS contributed equally to this work.

## Author Information

The authors declare no competing interests. Correspondence and requests for materials should be addressed to Alexander B. Khanikaev.

# Supplemental materials for *Experimental observation of topological exciton-polaritons in transition metal dichalcogenide monolayers*

Mengyao Li[1,2,3*], Ivan Sinev[4*], Fedor Benimetskiy[4], Tatyana Ivanova[4], Ekaterina Khestanova[4], Svetlana Kiriushechkina[1], Anton Vakulenko[1], Sriram Guddala[1,2], Maurice Skolnick[4,5], Vinod Menon[2,3], Dmitry Krizhanovskii[4,5], Andrea Alù[6,3,1], Anton Samusev[4], Alexander B. Khanikaev[1,2,3]

[1]Department of Electrical Engineering, City College of New York, New York, NY, USA

[2]Physics Department, City College of New York, New York, NY, USA

[3]Physics Program, Graduate Center of the City University of New York, New York, NY, USA

[4]Department of Physics and Engineering, ITMO University, Saint Petersburg, Russia

[5]Department of Physics and Astronomy, University of Sheffield, Sheffield S3 7RH, UK

[6]Photonics Initiative, Advanced Science Research Center, City University of New York, New York, NY, USA

*These authors contributed equally to the present work

## A. Tight-binding model calculation

The breathing (expanded or shrunken) honeycomb lattice has 6 sites in a single unit cell and is described (in momentum Bloch space) by the following Hamiltonian, which follows directly from the tight binding model [1]:

$$H_0 = \begin{pmatrix} \omega_0 & -\kappa & 0 & -je^{-ik_x} & 0 & -\kappa \\ -\kappa & \omega_0 & -\kappa & 0 & -je^{i(-\frac{1}{2}k_x+\frac{\sqrt{3}}{2}k_y)} & 0 \\ 0 & -\kappa & \omega_0 & -\kappa & 0 & -je^{i(\frac{1}{2}k_x+\frac{\sqrt{3}}{2}k_y)} \\ -je^{ik_x} & 0 & -\kappa & \omega_0 & -\kappa & 0 \\ 0 & -je^{i(\frac{1}{2}k_x-\frac{\sqrt{3}}{2}k_y)} & 0 & -\kappa & \omega_0 & -\kappa \\ -\kappa & 0 & -je^{-i(\frac{1}{2}k_x+\frac{\sqrt{3}}{2}k_y)} & 0 & -\kappa & \omega_0 \end{pmatrix}, \quad \text{(A1)}$$

where $\omega_0$ is on-site energy (frequency), which is set to be zero in what follows, $\kappa$ and $j$ are intracell and intercell coupling (hopping) coefficients, respectively, and $k_x$ and $k_y$ are the components of the dimensionless (normalized by the lattice constant $a_0$) Bloch vectors.

For simplicity, in what follows we apply k-p approximation near the Γ-point, where we introduce a unitary transformation operator

$$U_{6\times 6} = \frac{1}{\sqrt{6}} \begin{pmatrix} -1 & 1 & -1 & 1 & -1 & 1 \\ 1 & 1 & 1 & 1 & 1 & 1 \\ 1 & e^{i\pi/3} & e^{2i\pi/3} & -1 & e^{-2i\pi/3} & e^{-i\pi/3} \\ 1 & e^{2i\pi/3} & e^{-2i\pi/3} & 1 & e^{2i\pi/3} & e^{-2i\pi/3} \\ 1 & e^{-i\pi/3} & e^{-2i\pi/3} & -1 & e^{2i\pi/3} & e^{i\pi/3} \\ 1 & e^{-2i\pi/3} & e^{2i\pi/3} & 1 & e^{-2i\pi/3} & e^{2i\pi/3} \end{pmatrix}, \quad \text{(A2)}$$

This operator transforms the original Haimltonian breathing honeycomb lattice $H_{cir} = U_{6\times6} H_0 U_{6\times6}^{-1}$, and diagonalizes it at the Γ point with the left and right circularly polarized eigenstates $p_\pm$ and $d_\pm$, where the p and d letters correspond to the orbital momentum of photon of $l = 1$ (dipole) and $l = 2$ (quadrupole), respectively, and the subscript indicates handedness (spin $s = \pm 1$) of the modes.

We then expand the Hilbert space to span the excitonic degrees of freedom with the orbital momentum $m = \pm 1$ at K/K' valleys of the TMD monolayer, respectively. To this aim the Hamiltonian dimensions are increases by 2 (from 6x6 to 8x8) by adding two exciton states with energies $\omega_+ \equiv \omega_L$, $\omega_- \equiv \omega_R$, whose coupling to photonic degrees of freedom is characterized by the coefficients $q_{p\pm}$, $q_{d\pm}$. Here for simplicity we consider only the case of coupling to the photonic band of interest. We also note that the exciton dispersion can be neglected due the small values of photon wavenumber compared to the exciton wavenumber, and the excitons are always considered as having momentum close to K and K' point in the TMD.

In the case we are discussing below the time-reversal symmetry is preserved, and, therefore, these two excitonic states are always degenerate, $\omega_L = \omega_R = \omega$, and $q_{p(d)+} = q_{p(d)-} = q_{p(d)}$. We thus get an expression of non-dispersive excitonic Hamiltonian ("flat bands"):

$$H_{ex} = \begin{pmatrix} \omega_L & 0 \\ 0 & \omega_R \end{pmatrix} = \omega \begin{pmatrix} 1 & 0 \\ 0 & 1 \end{pmatrix}. \tag{A3}$$

It is known that these six photonic bands contain six different kind of modes: two singular modes, two dipole modes and two quadrupole modes.

$$H_{8\times8}^c = \begin{pmatrix} H_{cir} & Q \\ Q^\dagger & H_{ex} \end{pmatrix}, \tag{A4}$$

$$Q^\dagger = \begin{pmatrix} 0 & 0 & q_p & q_d & 0 & 0 \\ 0 & 0 & 0 & 0 & q_p & q_d \end{pmatrix}, \tag{A5}$$

where the form of $Q$ is dictated by the conservation of the angular momentum $s = m$, that is the photonic of the left $s = 1$ (right $s = -1$) helicity interact only with the excitons at K valley with $m = 1$ (K' valley with $m = -1$) due to the valley polarization (also see Section D below).

The respective transformation matrix then assumes the following form in the extended Hilbert space:

$$U_{8\times8} = \begin{pmatrix} U_{6\times6} & 0 \\ 0 & I_2 \end{pmatrix}, \tag{A6}$$

Then, in order to reveal the structure of excitons-photon interactions in the original basis, we perform a reverse transformation back to original basis

$$H_{8\times8} = U_{8\times8}^{-1} H_{8\times8}^c U_{8\times8} = \begin{pmatrix} H_0 & Q_c \\ Q_c^\dagger & H_{ex} \end{pmatrix}, \tag{A7}$$

and obtain

$$Q_c^{\dagger} = \frac{1}{\sqrt{6}} * \begin{pmatrix} q_p + q_d & q_1 + \frac{q_2^2}{q_d} & \frac{q_1^2}{q_p} - q_2 & -q_p + q_d & -q_1 + \frac{q_2^2}{q_d} & -\frac{q_1^2}{q_p} - q_2 \\ q_p + q_d & -\frac{q_1^2}{q_p} - q_2 & -q_1 + \frac{q_2^2}{q_d} & -q_p + q_d & \frac{q_1^2}{q_p} - q_2 & q_1 + \frac{q_2^2}{q_d} \end{pmatrix} \text{(A8)}$$

where $q_1 = (-1)^{\frac{1}{3}} q_p$ and $q_2 = (-1)^{\frac{1}{3}} q_d$. This gives us the form of coupling between the polaritonic bands and photonic bands with this $Q_c$ coupling block.

By introducing interaction of polaritonic bands with the photonic bands in the Dirac cone (without shrinking or expansion) states $p_{\pm}$ and $d_{\pm}$, by placing them both mid-gap and to cross with the upper bands, and we calculate the band structure of the resultant system. The respective results are given in Fig. S1 and Fig. S2.

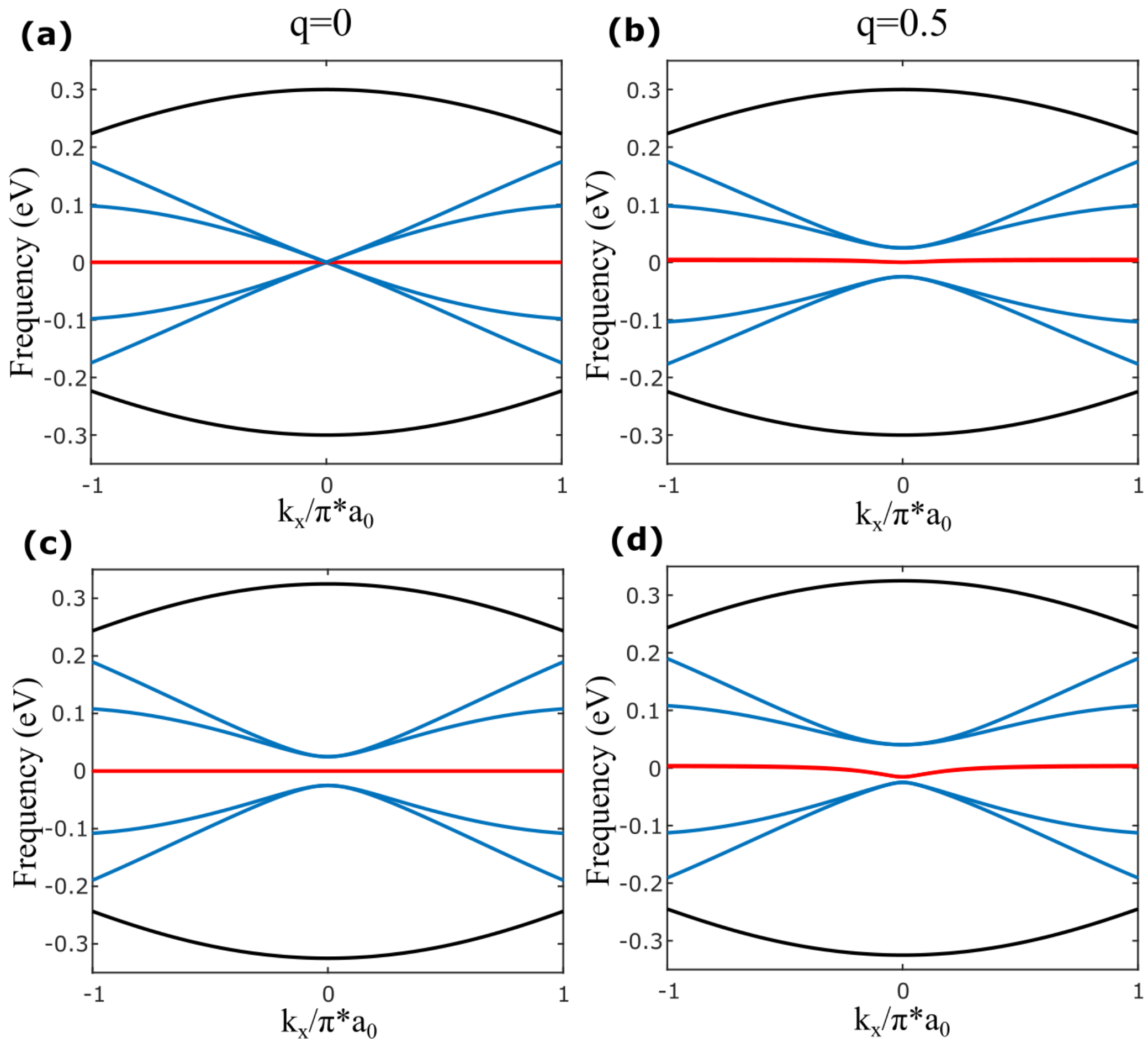

**Fig. S1: Band structure of infinite lattice with polaritons obtained from TBM model for exciton frequencies exactly at the Dirac point (mid-gap for shrunk expanded lattice). a** and **b**, unperturbed honeycomb lattice interacting with excitonic bands, q=0 (no interaction) and q=0.5 (with interaction) cases. **c** and **d**, topological (expanded) honeycomb lattice with excitonic states, q=0 (no interaction) and q=0.5 (with interaction) cases.

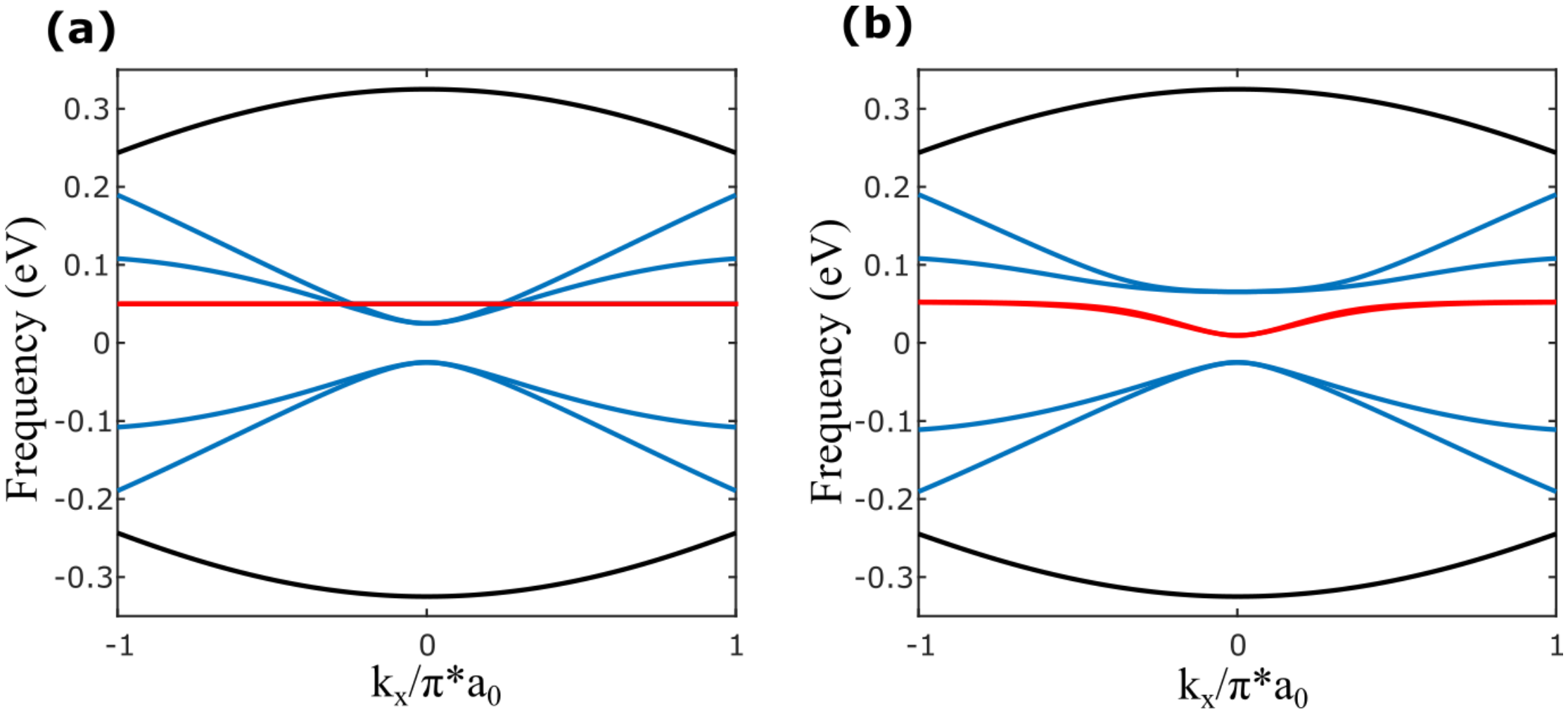

**Fig. S2: Band structure of infinite lattice with polaritons obtained from TBM for exciton frequencies crossing the upper photonic bands**. **a** and **b**, topological (expanded) photonic honeycomb lattice with excitonic states, for q=0 (no interaction) and q=0.5 (with interaction), respectively.

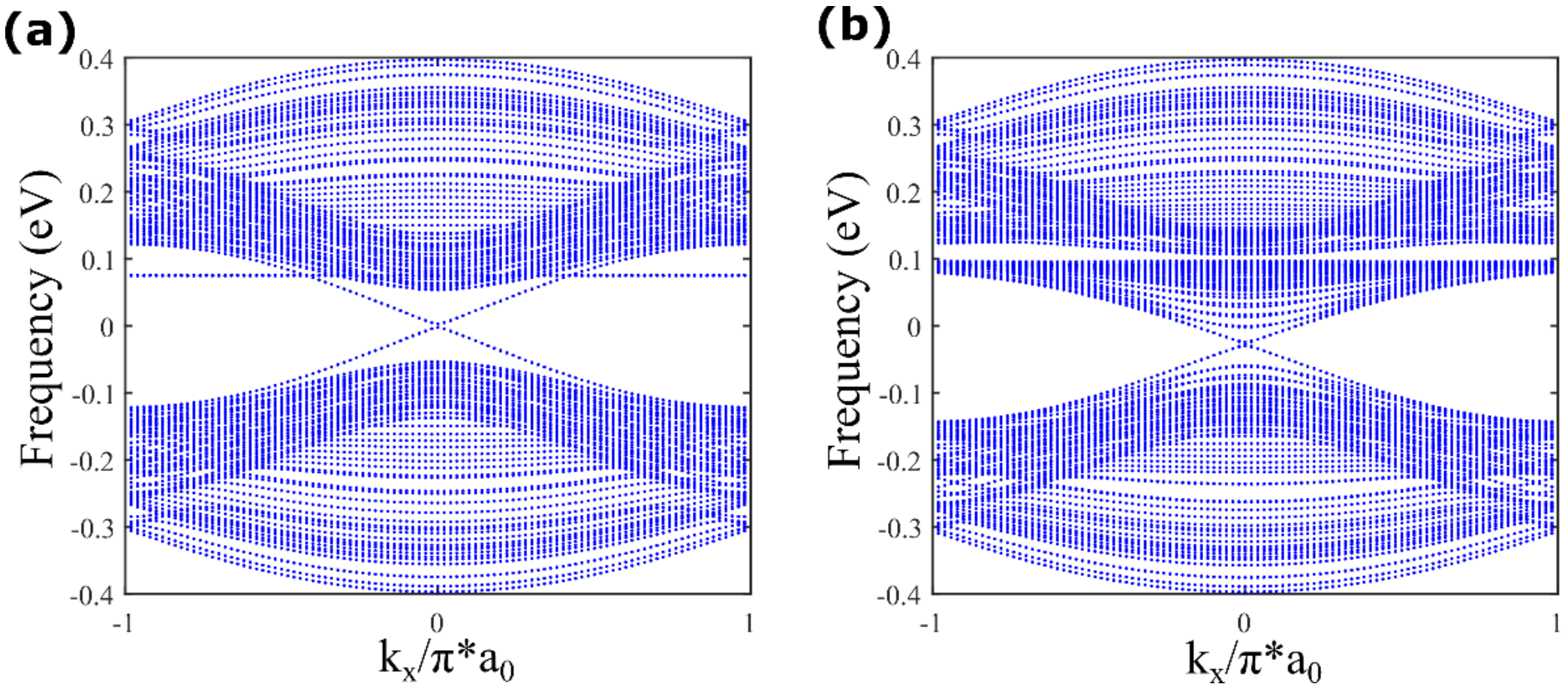

**Fig. S3: Band structure of a supercell with 10 trivial and 10 topological unit cells with the domain wall in the middle obtained from TBM for exciton frequencies crossing the upper photonic bands**. **a**, no coupling between exciton and photonic bands ($q = 0$), **b**, a strong coupling between polariton bands and photonic bands $q = 1.5$, which is 60% of the average coupling strength between sites $(\kappa + j)/2$.

To observe the emergence of the edge states, we also preformed calculations for the honeycomb supercell consisting of 10 shrunk (trivial) and 10 expanded (topological) unit cells, with the domain wall in the middle. The results in Fig. S3 clearly show that the edge states are transferred from photonic bandgap to the gap between the lower photonic bands and the new band corresponding to the exciton-polaritons, while there's no edge state at the upper gap, indicating the transfer of topological invariant from the former photonic bulk band to the polaritonic band. This is indeed confirmed by the direct calculation of the spin-Chern numbers for the new polaritonic bands, which yields $C_s = \pm 1$ for the two bands.

### B. Photonic treatment: plane wave expansion near the Γ point and the Berry curvature and spin-Chern number calculations

Without distortion (shrinkage or expansion), the band structure forms two Dirac cones at K and K′. After the distortion is introduced, the Brillouin zone folding occurs, and four-fold degeneracy at the Γ point is lifted and a gap opens. Following Ref. 1, we performed a plane-wave expansion around Γ point and reduced the dimension of resultant matrix to 6x6 by getting rid of the two singlet bands, and then further reduced the dimension of this matric to 3 by 3 matrix by focusing on a particular block corresponding to $s = 1$.

In our system corresponding to experiment, electric field is in-plane, therefore we chose to work with $H_z$ component. The Helmholtz equation for the respective photonic crystal then assumes the following form

$$\nabla \times \left(\frac{1}{\varepsilon(x,y)} \nabla \times\right) H_z + k_0^2 H_z = 0 \quad \text{(B1)}$$

where $k_0 = \frac{\omega}{c}$. In the following equations we replaced $k_0$ with parameter $p$.

$$H_z(\mathbf{r}) = \sum_G H_G \, e^{i(\mathbf{G}+\mathbf{k})\cdot\mathbf{r}} \quad \text{(B2)}$$

$$\frac{1}{\varepsilon}(\mathbf{r}) = \sum_G \kappa_G \, e^{i\mathbf{G}\cdot\mathbf{r}} \quad \text{(B3)}$$

$$k_0^2 H_G - \sum_{G'} \kappa_{G-G'} \left[(G_x + k_x)(G'_x + k_x) + (G_y + k_y)(G'_y + k_y)\right] H_{G'} = 0 \quad \text{(B4)}$$

By the integration $\kappa_G = \frac{1}{S_0} \int \frac{1}{\varepsilon(r)} e^{-i\mathbf{G}\cdot\mathbf{r}} \, d\boldsymbol{r}$, we can obtain all orders of refractive index Fourier components needed, $S_0$ is the unit cell area. $G = \frac{4\pi}{\sqrt{3}a}$ is the length of reciprocal lattice vectors, and we dropped the subscript z for the magnetic field for simplicity as it is redundant.

$$[\kappa_0 \, k^2 - p^2] H_0 + \kappa_1 [k^2 + G \, \mathrm{k_y}] H_1 + \kappa_1 \left[k^2 + G\left(\frac{\sqrt{3}}{2}\mathrm{k_x} + \frac{1}{2}\mathrm{k_y}\right)\right] H_2 + \kappa_1 \left[k^2 + G\left(\frac{\sqrt{3}}{2}\mathrm{k_x} - \frac{1}{2}\mathrm{k_y}\right)\right] H_3 + \kappa_1 (k^2 - G \, \mathrm{k_y}) H_4 + \kappa_1 \left[k^2 + G\left(-\frac{\sqrt{3}}{2}\mathrm{k_x} - \frac{1}{2}\mathrm{k_y}\right)\right] H_5 + \kappa_1 \left[k^2 + G\left(-\frac{\sqrt{3}}{2}\mathrm{k_x} + \frac{1}{2}\mathrm{k_y}\right)\right] H_6 = 0 \quad \text{(B5.1)}$$

$$\kappa_1 (k^2 + G\mathrm{k_y}) H_0 + [-p^2 + \kappa_0 (k^2 + 2G\mathrm{k_y} + G^2)] H_1 + \kappa_1 \left[k^2 + G\left(\frac{\sqrt{3}}{2}\mathrm{k_x} + \frac{3}{2}\mathrm{k_y}\right) + \frac{1}{2}G^2\right] H_2 + \kappa_2 \left(k^2 + G\left(\frac{\sqrt{3}}{2}\mathrm{k_x} + \frac{1}{2}\mathrm{k_y}\right) - \frac{1}{2}G^2\right) H_3 + \kappa_3 (k^2 - G^2) H_4 + \kappa_2 \left(k^2 + G\left(-\frac{\sqrt{3}}{2}\mathrm{k_x} + \frac{1}{2}\mathrm{k_y}\right) - \frac{1}{2}G^2\right) H_5 + \kappa_1 \left(k^2 + G\left(-\frac{\sqrt{3}}{2}\mathrm{k_x} + \frac{3}{2}\mathrm{k_y}\right) + \frac{1}{2}G^2\right) H_6 = 0 \quad \text{(B5.2)}$$

$\kappa_1\left(k^2+G\left(\frac{\sqrt{3}}{2}\mathrm{k_x}+\frac{1}{2}\mathrm{k_y}\right)\right)H_0+\kappa_1\left(k^2+G\left(\frac{\sqrt{3}}{2}\mathrm{k_x}+\frac{3}{2}\mathrm{k_y}\right)+\frac{1}{2}G^2\right)H_1+\left[-p^2+\kappa_0\left(k^2+2G\left(\frac{\sqrt{3}}{2}\mathrm{k_x}+\frac{1}{2}\mathrm{k_y}\right)+G^2\right)\right]H_2+\kappa_1\left(k^2+\sqrt{3}G\mathrm{k_x}+\frac{1}{2}G^2\right)H_3+\kappa_2\left(k^2+G\left(\frac{\sqrt{3}}{2}\mathrm{k_x}-\frac{1}{2}\mathrm{k_y}\right)-\frac{1}{2}G^2\right)H_4+\kappa_3(k^2-G^2)H_5+\kappa_2\left(k^2+G\mathrm{k_y}-\frac{1}{2}G^2\right)H_6=0$ (B5.3)

$\kappa_1\left(k^2+G\left(\frac{\sqrt{3}}{2}\mathrm{k_x}-\frac{1}{2}\mathrm{k_y}\right)\right)H_0+\kappa_2\left(k^2+G\left(\frac{\sqrt{3}}{2}\mathrm{k_x}+\frac{1}{2}\mathrm{k_y}\right)-\frac{1}{2}G^2\right)H_1+\kappa_1\left(k^2+\sqrt{3}\,G\,\mathrm{k_x}+\frac{1}{2}G^2\right)H_2+\left[-p^2+\kappa_0\left(k^2+2G\left(\frac{\sqrt{3}}{2}\mathrm{k_x}-\frac{1}{2}\mathrm{k_y}\right)+G^2\right)\right]H_3+\kappa_1\left(k^2+G\left(\frac{\sqrt{3}}{2}\mathrm{k_x}-\frac{3}{2}\mathrm{k_y}\right)+\frac{1}{2}G^2\right)H_4+\kappa_2\left(k^2-G\mathrm{k_y}-\frac{1}{2}G^2\right)H_5+\kappa_3(k^2-G^2)H_6=0$ (B5.4)

$\kappa_1(k^2-G\mathrm{ky})H_0+\kappa_3(k^2-G^2)H_1+\kappa_2\left(k^2+G\left(\frac{\sqrt{3}}{2}\mathrm{k_x}-\frac{1}{2}\mathrm{k_y}\right)-\frac{1}{2}G^2\right)H_2+\kappa_1\left(k^2+G\left(\frac{\sqrt{3}}{2}\mathrm{k_x}-\frac{3}{2}\mathrm{k_y}\right)+\frac{1}{2}G^2\right)H_3+\left[-p^2+\kappa_0(k^2-2G\mathrm{k_y}+G^2)\right]H_4+\kappa_1\left(k^2+G\left(-\frac{\sqrt{3}}{2}\mathrm{k_x}-\frac{3}{2}\mathrm{k_y}\right)+\frac{1}{2}G^2\right)H_5+\kappa_2\left(k^2+G\left(-\frac{\sqrt{3}}{2}\mathrm{k_x}-\frac{1}{2}\mathrm{k_y}\right)-\frac{1}{2}G^2\right)H_6=0$ (B5.5)

$\kappa_1\left(k^2+G\left(-\frac{\sqrt{3}}{2}\mathrm{k_x}-\frac{1}{2}\mathrm{k_y}\right)\right)H_0+\kappa_2\left(k^2+G\left(-\frac{\sqrt{3}}{2}\mathrm{k_x}+\frac{1}{2}\mathrm{k_y}\right)-\frac{1}{2}G^2\right)H_1+\kappa_3(k^2-G^2)H_2+\kappa_2\left(k^2-G\mathrm{k_y}-\frac{1}{2}G^2\right)H_3+\kappa_1\left[k^2+G\left(-\frac{\sqrt{3}}{2}\mathrm{k_x}-\frac{3}{2}\mathrm{k_y}\right)+\frac{1}{2}G^2\right]H_4+\left[-p^2+\kappa_0\left(k^2+2G\left(-\frac{\sqrt{3}}{2}\mathrm{k_x}-\frac{1}{2}\mathrm{k_y}\right)+G^2\right)\right]H_5+\kappa_1\left(k^2-\sqrt{3}G\,\mathrm{k_x}+\frac{1}{2}G^2\right)H_6=0$ (B5.6)

$\kappa_1\left(k^2+G\left(-\frac{\sqrt{3}}{2}\mathrm{k_x}+\frac{1}{2}\mathrm{k_y}\right)\right)H_0+\kappa_1\left(k^2+G\left(-\frac{\sqrt{3}}{2}\mathrm{k_x}+\frac{3}{2}\mathrm{k_y}\right)+\frac{1}{2}G^2\right)H_1+\kappa_2\left(k^2+G\mathrm{k_y}-\frac{1}{2}G^2\right)H_2+\kappa_3(k^2-G^2)H_3+\kappa_2\left(k^2+G\left(-\frac{\sqrt{3}}{2}\mathrm{k_x}-\frac{1}{2}\mathrm{k_y}\right)-\frac{1}{2}G^2\right)H_4+\kappa_1\left(k^2-G\sqrt{3}\mathrm{k_x}+\frac{1}{2}G^2\right)H_5+\left[-p^2+\kappa_0\left(k^2+2G\left(-\frac{\sqrt{3}}{2}\mathrm{k_x}+\frac{1}{2}\mathrm{k_y}\right)+G^2\right)\right]H_6=0$ (B5.7).

These seven equations establish an effective Hamiltonian for our system.

We then performed a unitary transformation,

$$\hat{U}_{PW}=\frac{1}{\sqrt{3}}\begin{pmatrix}\sqrt{3}&0&0&0&0&0&0\\0&0&1&0&1&0&1\\0&1&0&1&0&1&0\\0&0&1&0&\mathrm{Exp}[4\pi\,i/3]&0&\mathrm{Exp}[2\pi\,i/3]\\0&0&1&0&\mathrm{Exp}[2\pi\,i/3]&0&\mathrm{Exp}[4\pi\,i/3]\\0&1&0&\mathrm{Exp}[4\pi\,i/3]&0&\mathrm{Exp}[2\pi\,i/3]&0\\0&1&0&\mathrm{Exp}[2\pi\,i/3]&0&\mathrm{Exp}[4\pi\,i/3]&0\end{pmatrix},\text{ (B6)}$$

which block diagonalized this matrix into 3x3 and 4x4 blocks at Gamma point

$$\hat{H}_{3\times3} = \begin{pmatrix} -p^2 & 0 & 0 \\ 0 & -p^2 + G^2(\kappa_0 - \kappa_2) & G^2(\kappa_1 - \kappa_3) \\ 0 & G^2(\kappa_1 - \kappa_3) & -p^2 + G^2(\kappa_0 - \kappa_2) \end{pmatrix}, \quad \text{(B7)}$$

$$\hat{H}_{4\times4} = \begin{pmatrix} h_{1a} & 0 & h_{1b} & 0 \\ 0 & h_{1a} & 0 & h_{1b}^* \\ h_{1b}^* & 0 & h_{1a} & 0 \\ 0 & h_{1b} & 0 & h_{1a} \end{pmatrix}, \quad \text{(B8)}$$

where $h_{1a} = -p^2 + G^2(\kappa_0 + \kappa_2/2)$ and $h_{1b} = (1 + i\sqrt{3})G^2(\kappa_1/2 + \kappa_3)$. The 4x4 block $H_{4\times4}$ corresponds to the 4 bands forming the Dirac cone.

Addition of k-dependent term obtained from the $k$ dot $p$ theory yield the effective Hamiltonian near Gamma point of the following form (more detail can be found in Supplement, section 2, of Ref. [1])

$$\hat{H}_k = B_2|\mathbf{k}|^2\hat{I} + \begin{pmatrix} u_1 - B(k_x^2 + k_y^2) & 2G(-ik_x + k_y) & 0 & 0 \\ 2G(ik_x + k_y) & u_2 + B(k_x^2 + k_y^2) & 0 & 0 \\ 0 & 0 & u_1 - B(k_x^2 + k_y^2) & 2G(-ik_x - k_y) \\ 0 & 0 & 2G(ik_x - k_y) & u_2 + B(k_x^2 + k_y^2) \end{pmatrix} \quad \text{(B9)}$$

$$u_1 = G^2(\kappa_1 + 2\kappa_3 + 2\kappa_0 + \kappa_2) - 2p^2$$

$$u_2 = G^2(-\kappa_1 - 2\kappa_3 + 2\kappa_0 + \kappa_2) - 2p^2$$

$$B = 2(\kappa_3 - \kappa_1)$$

$$B_2 = 2(\kappa_0 - \kappa_2)$$

This expression is analogous to the Hamiltonian of the BHZ model. We then shift the Dirac cone to zero-energy, rewriting the Hamiltonian in the BHZ form:

$$\hat{H}_k = B_2|\mathbf{k}|^2\hat{I} + \begin{pmatrix} M - B|\mathbf{k}|^2 & A(-ik_x + k_y) & 0 & 0 \\ A(ik_x + k_y) & -M + B|\mathbf{k}|^2 & 0 & 0 \\ 0 & 0 & M - B|\mathbf{k}|^2 & A(-ik_x - k_y) \\ 0 & 0 & A(ik_x - k_y) & -M + B|\mathbf{k}|^2 \end{pmatrix}, \quad \text{(B10)}$$

where $M = (u_1 - u_2)/2$ , $A = 2G$, and $|\mathbf{k}|^2 = k_x^2 + k_y^2$ .

Then, as in Section A above, we expand the Hamiltonian to include excitons and introduce exciton-photon interactions

$$\hat{H}_D^c = \begin{pmatrix} M-B(k_x^2+k_y^2) & A(-ik_x+k_y) & 0 & 0 & q_p & 0 \\ A(ik_x+k_y) & -M+B(k_x^2+k_y^2) & 0 & 0 & q_d & 0 \\ 0 & 0 & M-B(k_x^2+k_y^2) & A(-ik_x-k_y) & 0 & q_p \\ 0 & 0 & A(ik_x-k_y) & -M+B(k_x^2+k_y^2) & 0 & q_d \\ q_p & q_d & 0 & 0 & \omega_L & 0 \\ 0 & 0 & q_p & q_d & 0 & \omega_R \end{pmatrix}$$

(B11).

The bands of $\hat{H}_D^c$ are spin-degenerate forming two groups of Dirac cone on top of each other spectrally, therefore we can further focus on the spin-up block establishing the 3x3 Hamiltonian for the respective spin:

$$\hat{H}_D' = \begin{pmatrix} M-B(k_x^2+k_y^2) & A(-ik_x+k_y) & q_p \\ A(ik_x+k_y) & -M+B(k_x^2+k_y^2) & q_d \\ q_p & q_d & \omega_L \end{pmatrix} \quad \text{(B12)}$$

We also calculated the Berry curvature and calculated the spin-Chern numbers of each bands of the Hamiltonian. When $u_1 > u_2$ and $B > 0$, or $u_1 < u_2$ and $B < 0$, the system is topological, otherwise it is trivial, yielding zero spin-Chern numbers on any band. We again see that the nontrivial spin-Chern number transfers to the polariton bands. Thus, the spin-Chern number for spin-up states without coupling are {1,0,-1} when the polariton frequency is at midgap and not crossing any band, while after applying a strong coupling with the upper photonic bands, the spin-Chern number for spin-up states transform into {0,1,-1}, and we can see that the nonzero topology invariant has indeed been transferred from upper photonic band to the polaritonic band.

## C. Circular wave excitation in TBM

To emulate the experimental conditions, a model of coupled mode theory with excitation source in Tight-Binding Model is introduced. In experiment, a circular polarized source is applied on the whole structure, and measurements of field took place on the boundary of topological and trivial domains, the EM field of a few boundary unit cells are collected.

In TBM model introduced in Section A (Eq.A1,A7,A8), the Hamiltonian of a periodic lattice with excitons $H_{8\times 8}$ contains 8 degrees of freedom, 6 of which are photonic and the other 2 are excitonic. We apply excitation source $S_{in}$ on these 6 photonic bands.

$$S_{in} = \begin{pmatrix} 1 & e^{\pm\frac{i\pi}{3}} & e^{\pm\frac{2i\pi}{3}} & e^{\pm i\pi} & e^{\pm\frac{4i\pi}{3}} & e^{\pm\frac{5i\pi}{3}} & 0 & 0 \end{pmatrix}^{\mathsf{T}}, \quad \text{(C1)}$$

Where plus and minus indicates different field rotation directions, or directions of circular polarizations of light source.

The wavefunction of the system ψ have this relation under excitation

$$\frac{1}{i}\frac{d\psi}{dt} = (\widehat{H}_{8\times 8} + i\hat{\gamma})\psi + \alpha S_{in}, \quad \text{(C2)}$$

Where α is the coupling coefficient of the excitation source, and $\hat{\gamma}$ describes is the total loss of the modes of the system. By solving Eq. C2 we can get energy distribution $|\psi|^2$ and reflectivity $r = S_{out} = \alpha\psi$, $R = \alpha^2|\psi|^2$ of the system. Moreover, we can also separate the photonic part and excitonic part in ψ by corresponding index (e.g. in this case, the first 6 elements in ψ is photonic, and the last 2 are excitonic, corresponding to Hamiltonian $H_{8\times 8}$), and the contribution of photonic and excitonic part of the bulk or edge spectrum can thus be shown separately. This method can also be used in supercell Hamiltonian in TBM, where the structure is periodic in one direction, the domain boundaries appeared in the other direction, and by calculating such structure we can get edge state spectrum of the structure. In Fig.4 of the main text, we estimated the coupling coefficient as well as loss parameters γ and α for photonic and excitonic bands by fitting with frequencies, gap widths and linewidths in experimental results, and obtained the photonic and excitonic edge state spectra for both cases with different relative position of excitonic bands corresponding to experiments.

**D. Exciton modes coupling evaluation**

It's known that excitons in monolayer TMDs are doubly degenerate modes with angular momentum $m = \pm 1$ which represent pairs of electrons and holes at K and K' point, in the atomic Brillion zone, respectively. Consisting of a pair of charged particles, these excitonic states have nonzero dipole moment and can be described as the dipolar modes two dipolar modes of opposite helicity. The photonic metasurface in hands, however, possesses more rich structure, with four photonic bands in the frequency range of interest, corresponding to two dipolar $p_\pm$ $(l = 1)$ modes and two quadrupolar $d_\pm$ $(l = 2)$ modes of opposite helicities $(s = \pm 1)$. It is intuitive to think that the dipolar exitonic modes will couple to the dipolar photonic modes, respectively. However, whether the coupling would happen between exitonic and quadrupolar $d_\pm$ modes is not as straightforward because of a completely different spatial scales on which variation of photonic and excitonic wavefunctions occurs, and therefore, it should be carefully analyzed. In this section we aim at finding the coupling between modes various photonic modes and excitonic modes.

The response of a TMD monolayer is homogeneous on the scale of variation of photonic field, and, therefore, it can be expressed by dielectric constant, or, equivalently, by the following surface (sheet) conductivities:

$$\hat{\sigma}_+ = \frac{\sigma_{TMD}}{2}\begin{pmatrix} 1 & -i \\ i & 1 \end{pmatrix}, \hat{\sigma}_- = \frac{\sigma_{TMD}}{2}\begin{pmatrix} 1 & i \\ -i & 1 \end{pmatrix}, \quad \text{(D1)}$$

where $\sigma_{TMD}$ is the total high-frequency (optical-frequency) surface conductivity of a monolayer, and $\hat{\sigma}_+$ and $\hat{\sigma}_-$ are contributions to the conductivity from K-valley excitons $(m = +1)$ and K'-valley excitons $(m = -1)$, respectively. The structure of these tensors is dictates by the selection rule on exciton-photon interaction due to the angular momentum conservation, which ensures that the spin of photon should equal the angular momentum of excitons $(s = m)$. It is also clear that the cumulative response of the TMD monolayer $\hat{\sigma}_{TMD} = \hat{\sigma}_+ + \hat{\sigma}_- = \sigma_{TMD}$, i.e. isotropic, unless time-reverals symmetry is broken ($\hat{\sigma}_+ \neq \hat{\sigma}_-$).

Note that, because the conductivity tensors $\hat{\sigma}_\pm$ do not vary in space, TMD does not mix dipolar and quadrupolar photonic states since they are orthogonal.

We start with the dipolar photonic bands $l = 1$ , $s = \pm 1$, which have the following form:

$$|\boldsymbol{E}_{p+}\rangle = \frac{f(r)}{\sqrt{2}}\begin{pmatrix}1\\ i\end{pmatrix}, \ |\boldsymbol{E}_{p-}\rangle = \frac{f(r)}{\sqrt{2}}\begin{pmatrix}1\\ -i\end{pmatrix}, \tag{D2}$$

where $f(\boldsymbol{r})$ describes the field distribution within the photonic unit cell, and $f(\boldsymbol{r}) \approx 1$ in the weak crystal regime used in the plane wave expansion above. The field vectors are also normalized such that $\langle \boldsymbol{E}_{ps} | \boldsymbol{E}_{ps} \rangle = \int dV\, f(\boldsymbol{r}) = 1$, where integration is over the unit cell volume ($dV = dSdz$).

The interaction strength between the excitons ($m = \pm 1$) and the dipolar photonic states ($s = \pm 1$) can be then calculated as $q^2_{p(m,s)} = N \int dS_{TMD} \langle \boldsymbol{E}_{ps} | \boldsymbol{J}_{pm} \rangle = N \int dS_{TMD} \langle \boldsymbol{E}_{ps} | \hat{\sigma}_m | \boldsymbol{E}_{pm} \rangle$, where we used the expression for the excitonic current modes in the TMD monolayer $|\boldsymbol{J}_{pm}\rangle = \hat{\sigma}_m |\boldsymbol{E}_{pm}\rangle$ driven by the electric field $|\boldsymbol{E}_{pm}\rangle$, the integration is over the surface of 2D material in the unit cell, and $N$ is the normalization factor. Then, since the conductivity is homogeneous over the surface, we can readily obtain $q_{d(s,m)} = (N\langle E_s | \hat{\sigma}_m | E_m \rangle)^{\frac{1}{2}} = \sqrt{\frac{N\sigma_{TMD}}{2}}\,\delta_{sm}$. This confirms that the form of the TMD response chosen yields interactions with exciton-photon interactions which respects the conservation of the angular momentum $s = m$, and, consequently, the valley polarization.

Considering the photonic quadrupolar modes ($l = 2$), the field profiles have more elaborate form with the spatial distribution ($d_\pm = d_{x^2-y^2} \pm i d_{xy}$), where, as before, $\pm$ indicates the photon spin $s = \pm 1$, and the field profiles for the case of the weak modulation in the crystal assume the following form in the vector representation:

$$\boldsymbol{E}_{d+} \sim \begin{pmatrix} x + iy \\ -y + ix \end{pmatrix}, \ \boldsymbol{E}_{d-} \sim \begin{pmatrix} x - iy \\ -y - ix \end{pmatrix}. \tag{D3}$$

Then, the coupling of the excitons and quadrupolar photonic states can be found by evaluating the following expressions: $q^2_{d(m,s)} = N \int dS_{TMD} \langle \boldsymbol{E}_{ds} | \boldsymbol{J}_{dm} \rangle = N \int dS_{TMD} \langle \boldsymbol{E}_{ds} | \hat{\sigma}_m | \boldsymbol{E}_{dm} \rangle$, which can be readily evaluated to yield $q^2_{d(+,+)} = q^2_{d(-,-)} \sim \int 2(x^2 + y^2)\, ds \neq 0$, while $q_{d(+,-)} = q_{d(-,+)} = 0$, which again establishes a section rule on interactions between excitons and quadrupolar photons ensuring condition $s = m$. Therefore, the orbital degree of freedom $l$ does not play any significant role in defining selection rules of exciton-photon interactions, which can be interpreted as the result of the orders of magnitude difference in the length scales on which these quantities are defined. Note, however, that the orbital degree of freedom of light $l$ is crucial for the topological photonic phase stemming from the spin-orbit interactions of light.

As we can see here, the two conclusions we arrive at from considerations of this section: (*i*) there is the spin selection rules for exciton-photon coupling with both dipolar and quadrupolar photonic modes, and (ii) coupling of excitons with quadrupole modes is not equal to zero since the selection rule with respect to the orbital momentum of light does not apply.

## E. Effective phase winding in exciton-photon interactions

To demonstrate the emergence of phase winding in coupling between photonic and excitonic bands, here we investigate a massive Dirac Hamiltonian interacting with exciton state for one particular spin. The parabolic corrections to the energies (term B in BHZ Hamiltonian) is thus neglected ($B = 0$), which, however, does not affect the generality of our conclusions, since the parabolic correction is known to only change the Chern number by ½ [1], which, as our numerical calculations show, remains true for the more general cases of BHZ and tight-binding models. Thus, we investigate the spin-up ($s = +1$) subsystems of photonic system described by the Hamiltonian

$$\hat{H}^{+}_{2\times2} = \hat{\sigma}_x k_x + \hat{\sigma}_y k_y + M\hat{\sigma}_z = \begin{pmatrix} M & k_x - ik_y \\ k_x + ik_y & -M \end{pmatrix}, \qquad \text{(E1)}$$

where $\hat{\sigma}_n$ are Pauli matrices, and Dirac velocity was assumed to be $v_D = 1$ without any loss of generality. The two degrees of freedom in Eq. (E1) are implicitly associated with the angular momentum $l = 1$ and $l = 2$ for the case of our system

We can find the eigenmodes of this Hamiltonian for the eigenvalue problem $\hat{H}_+ \bar{\psi}^+_n = \omega_n \bar{\psi}^+_n$, which have the well-known form

$$\bar{\psi}_1 = \frac{1}{N_1}\begin{pmatrix} 1 \\ f_1 e^{i\theta} \end{pmatrix}, \bar{\psi}_2 = \frac{1}{N_2}\begin{pmatrix} 1 \\ f_2 e^{-i\theta} \end{pmatrix}, \qquad \text{(E2)}$$

and the spectrum $\omega_{1(2)} = \pm\sqrt{k^2 + M^2}$, where $k = |\boldsymbol{k}|$ and $\theta = \tan^{-1}\left(\frac{k_y}{k_x}\right)$, and $f_{1(2)} = -\frac{k}{m - \omega_{1(2)}}$, and $N_{1(2)} = |\bar{\psi}_{1(2)}|$ is the normalization factor.

Thus, the Hamiltonian (E1) can be diagonalized by the unitary transformation

$$\hat{U}_{2\times2} = [\bar{\psi}_1, \bar{\psi}_2]. \qquad \text{(E3)}$$

By expanding the Hamiltonian to include $m = +1$ exciton with the energy $\omega_{ex} = -M$ and exciton-photon interactions $q = q^*$ with the lower band, we obtain

$$\hat{H}^{+}_{3\times3} = \begin{pmatrix} M & k_x - ik_y & 0 \\ k_x + ik_y & -M & q \\ 0 & q & -M \end{pmatrix}, \qquad \text{(E4)}$$

which can be block diagonalized by applying the expanded unitary transformation

$$\hat{U}_{3\times3} = diag[\hat{U}, 1], \qquad \text{(E5)}$$

yielding

$$\hat{\hat{H}}^{+}_{3\times3} = \begin{pmatrix} \omega_1(k) & 0 & q/N_1 \\ 0 & \omega_2(k) & qf_2/N_2 e^{i\theta} \\ q/N_1 & qf_2/N_2 e^{-i\theta} & -M \end{pmatrix}, \qquad \text{(E6)}$$

Where one can clearly see the phase to emerge in the coupling between the (diagonalized) photonic eigenstate and the exciton. Then, we consider the fact that the exciton is degenerate at k=0 with the lower (2nd) band and use the degenerate perturbation theory. Therefore, interaction with the upper (1st) band can be neglected giving the effective Hamiltonian

$$H_{2\times2}^{+(eff)} = \begin{pmatrix} \omega_2(k) & \tilde{q}e^{i\theta} \\ \tilde{q}e^{-i\theta} & -M \end{pmatrix}, \tag{E7}$$

where $\tilde{q}(k) = qf_2/N_2$ is the strength of the effective exciton-photon coupling.

We consider the effect of such interactions near $k = 0$, where the degeneracy takes place, to understand the structure of the eigenmode and energy splitting. In this case the energy $\omega_2$ can be approximated with $\omega_2 = -M - \frac{k^2}{2M}$ . Shifting the global energy by $+M$, we then obtain the eigenvalue problem:

$$\begin{pmatrix} -\frac{k^2}{2M} & \tilde{q}e^{i\theta} \\ \tilde{q}e^{-i\theta} & 0 \end{pmatrix} \tilde{\psi} = \tilde{\omega}\tilde{\psi}, \tag{E8}$$

which, for $k\sim0$, gives eigenvalues $\tilde{\omega} \approx -\frac{k^2}{4M} \pm \tilde{q}$, indicating avoided crossing with the splitting of $2\tilde{q}$ near Γ-point, and both eigenstates $\tilde{\psi}\sim[1, \pm e^{-i\theta}]$become 50% excitonic and 50% photonic.

## F. Processing of the angle-resolved reflectivity maps

Since the samples were fabricated using SOI substrates, all the measured angle-resolved reflectivity maps possessed characteristic Fabry-Pérot background, see Fig. S4a. This feature hinders the observation of the modes of topolaritonic system. In order to get rid of this background without affecting the spectral position of the reflectivity features of interest, we resorted to a post-processing procedure. The slowly varying background in the measured map was approximated with a cubic smoothing spline, with a smoothing parameter chosen such that in the resulting differential reflectivity map the Fabry-Pérot pattern fully vanishes. This resulted in images with good visibility of the photonic and polaritonic modes of the structure, see Fig. S4b and reflectivity maps throughout the main manuscript.

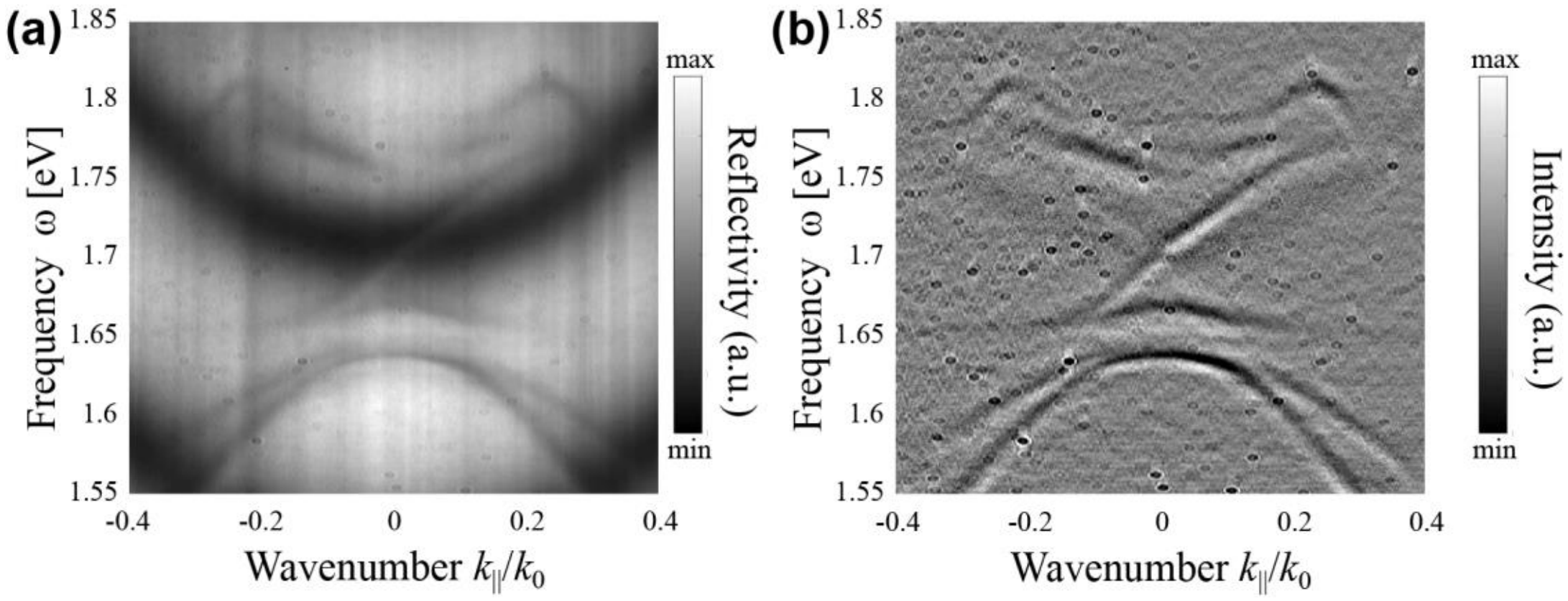

**Fig. S4:** **a**, raw data and **b**, post-processed angle-resolved reflectivity map of the sample shown in Fig. 4c of the main text.

## G. Experiments with a thicker hBN layer

Figure S5 shows the angle-resolved reflectance maps for the sample with 60-nm-thick hBN layer between the metasurface and the $MoSe_2$ monolayer. Red shift of the photonic modes driven by increased thickness of hBN leads to the case when the exciton frequency is inside the topological gap of the metasurface and do not exhibit strong coupling with the photonics modes.

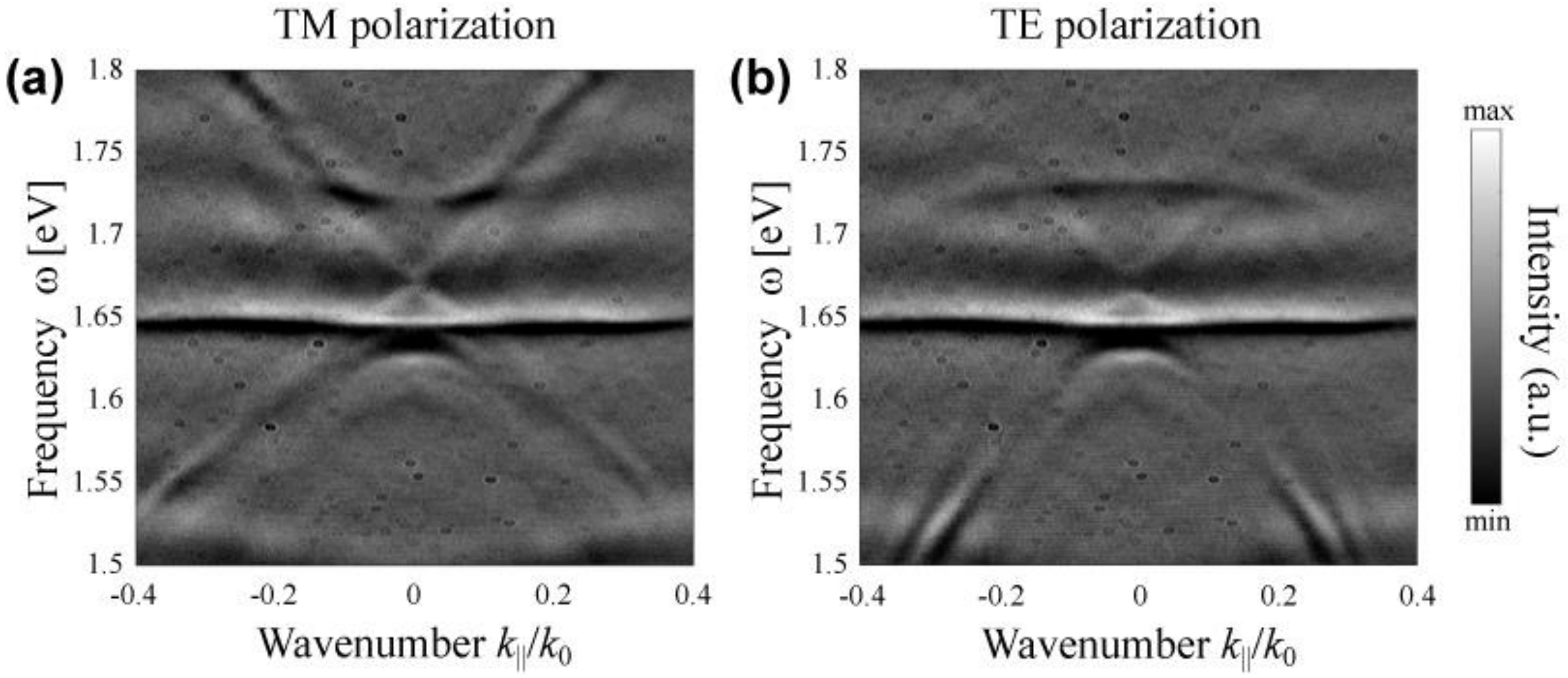

**Fig. S5: Angle-resolved reflectivity of sample with thick hBN layer**. **a**, TM polarized excitation **b**, TE-polarized excitation.